\documentstyle[12pt]{article}
\normalbaselines

\begin{document}
\begin{flushright}

{\raggedleft
SWAT/96/131\\

HUB-EP-96/56\\

hep-lat/9611005\\[1cm]}

\end{flushright}

\renewcommand{\thefootnote}{\fnsymbol{footnote}}
\begin{center}
{\LARGE\baselineskip0.9cm 
The exact equivalence of the one-flavour 

lattice Thirring model with Wilson fermions 

to a two-colour loop model\\[1.5cm]}

{\large K. Scharnhorst\footnote[2]{E-mail:
{\tt scharnh@linde.physik.hu-berlin.de}}
}\\[0.3cm]
{\small University of Wales Swansea

Department of Physics

Singleton Park

Swansea, SA2 8PP 

United Kingdom}\\[0.3cm]

{\small and}\\[0.3cm]

{\small Humboldt-Universit\"at zu Berlin\footnote[3]{Present address}

Institut f\"ur Physik

Invalidenstr.\ 110

D-10115 Berlin

Federal Republic of Germany}\\[1.5cm]

\begin {abstract}
Within Euclidean lattice field theory an exact equivalence
between the one-flavour 2D Thirring model with Wilson fermions
and Wilson parameter $r = 1$ to a two-colour loop model
on the square lattice is established. For non-interacting 
fermions this model reduces to an exactly solved 
loop model which is known to be a free fermion model.
The two-colour loop model equivalent to the 
Thirring model can also be understood as a 4-state 49-vertex model.
\end{abstract}

\end{center}

\renewcommand{\thefootnote}{\arabic{footnote}}

\thispagestyle{empty}

\newpage
\section{Introduction}

Four-fermion models on the lattice and in the continuum
(in various dimensions) have received a 
great deal of attention in recent years. This interest
is related to the problem of dynamical symmetry breaking
\cite{rose}, \cite{hand} as well as to the possible role of 
four-fermion interactions in QED in 4D \cite{leun1}, \cite{leun2}.
However, the reliable investigation of these models 
requires the application of non-perturbative and exact methods
which are hard to come by. Within lattice field theory
the investigation of fermionic
theories is hampered by the notorious difficulties encountered
in their numerical simulation, in particular, when the number of 
fermion species involved is odd. The development of new methods 
for this class of theories is therefore highly desirable.\\

Following a method developed by Salmhofer \cite{salm} in the 
exact study of the one-flavour strong-coupling lattice Schwinger model
with Wilson fermions, we have recently suggested \cite{scharn1}
that this method can also be applied to a wide class of purely 
fermionic theories with Wilson fermions in various dimensions. The study
of the one- and two-flavour strong-coupling Schwinger models
has lead to exact equivalences to statistical models 
which can be understood as either q-state vertex models or 
loop models \cite{salm}, \cite{scharn1}. In contrast to 
strongly coupled gauge theories with staggered fermions, which 
lead to pure monomer-dimer systems \cite{ross}, the 
consideration of Wilson fermions results in a different 
class of model, as these examples demonstrate.\\

In the present article, as a first step towards the application of the 
new method to four-fermion theories, within Euclidean lattice 
field theory we investigate the one-flavour 2D
Thirring model with Wilson fermions. In the past 
decades the one-flavour Thirring model in 2D has been widely 
studied in the continuum as well as within the Hamiltonian lattice 
approach. The massive Thirring model in the continuum is 
known to be quantum integrable (\cite{kore}, see also 
\cite{zhou}) and to be equivalent to the sine-Gordon model \cite{cole}. 
Within the Hamiltonian lattice approach 
equivalences of the Thirring model to other statistical models
have been found previously \cite{luth}, \cite{lusc} (see also
the references cited in \cite{zhou}). These models are related to 
staggered fermions. The only attempt so far to study the 
one-flavour Thirring model with multi-component spinors 
has been made by Ishida and Saito \cite{ishi}.
They however consider the case of naive fermions 
(`Wilson fermions with Wilson parameter $r = 0$')
only, which turns out to be much more involved than the 
case of Wilson fermions with $r = 1$ we are going 
to investigate in the present article.\\

The method we will employ is based on the fact that the 
Grassmann integration on the lattice can be performed exactly.
The Grassmann integration will consecutively be carried out
on the even and odd sublattices. In this process the possible 
contributions to the partition function will be analyzed
(for the method see also \cite{salm}, \cite{scharn1}).
In order to simplify the required algebraic manipulations with
Grassmann variables we make use of a Mathematica program \cite{math} 
specially designed for the present purpose. Since we consider only
one fermion species, throughout the paper we pay careful attention
to the ordering problem for the Grassmann variables involved
in order to keep track of any minus signs which possibly might
lead to negative contributions to the partition function.\\

The plan of the article is as follows. The main investigation 
is presented in section 2 which is divided into several subsections.
This division is introduced in order to give the reader easier 
access to the different steps in establishing the transformation from 
the one-flavour lattice Thirring model with Wilson
fermions to the equivalent statistical model. The structure of 
section 2 is explained in further detail at the end of its 
introductory part (before subsection 2.1 starts). Section 3
contains some discussion and conclusions. Finally, the proof 
of a mathematical result needed in subsection 2.2 is deferred 
to an Appendix.\\

\section{The loop model equivalence}

The partition function $Z_\Lambda$ of the one-flavour Thirring 
model with Wilson fermions on the square lattice $\Lambda$ is 
given by 

\parindent0.em

\begin{eqnarray}
\label{B1}
Z_\Lambda &=& \int D\psi D\bar\psi\ \ {\rm e}^{-S}\ \ \ ,
\end{eqnarray}

where $D\psi D\bar\psi = \prod_{x\in\Lambda}
d\psi_1(x)\ d\bar\psi_1(x)\ d\psi_2(x)\ d\bar\psi_2(x)$ 
denotes the multiple Grassmann integration on the lattice
(The subscripts are spinor indices.). The action $S$ is defined by

\begin{eqnarray}
\label{B2}
S &=& \sum_{x\in \Lambda} \left( {1\over 2}
\sum_\mu \left(\bar\psi(x+ e_\mu)(1+\gamma_\mu) \psi(x) +
\bar\psi(x)(1-\gamma_\mu) \psi(x+ e_\mu)
\right) \right.\nonumber\\[0.3cm]
&&\ \ \ -\ M \bar\psi(x)\psi(x) - {G\over 2}\ J_\mu (x) J_\mu (x)\Bigg)
\end{eqnarray}

with 

\begin{eqnarray}
\label{B3}
J_\mu (x) &=& \bar\psi(x) \gamma_\mu \psi(x)\ \ \ \ ,\\[0.3cm]
J_\mu (x) J_\mu (x) &=& 4\ \bar\psi_1(x)\bar\psi_2(x)
\psi_1(x)\psi_2(x)\ \ \ \ \ .
\end{eqnarray}

In eq.\ (\ref{B2}) the Wilson parameter $r$ has been set to 1; 
the hopping parameter $\kappa$ is consequently given by $\kappa = 1/2M$. 
Eq.\ (\ref{B1}) can be written as follows.

\begin{eqnarray}
\label{B4}
Z_\Lambda &=& \sum_{k_l \in \{0,1,2,3,4\}}
\ \int D\psi D\bar\psi\ \ 
\exp\left(\ \sum_{x\in \Lambda} M \bar\psi(x)\psi(x)\right)
\nonumber\\[0.3cm]
&&\ \ \ \times\ 
\exp\left(2 G\sum_{x\in \Lambda} \bar\psi_1(x)\bar\psi_2(x)
\psi_1(x)\psi_2(x)\right)
\nonumber\\[0.3cm]
&&\ \ \ \times\ \prod_{l = (x,x+e_\mu)} {1\over k_l !}
\left[ \bar\psi(x+ e_\mu) T_\mu^{(+)}\psi(x) +
\bar\psi(x) T_\mu^{(-)}\psi(x+ e_\mu) \right]^{k_l} .
\end{eqnarray}

The occupation numbers $k_l$
cannot exceed 4 because on each lattice site only 4 Grassmann variables
are present. Explicit calculation however shows that by virtue
of the nilpotency of the Grassmann variables $k_l$ cannot exceed 2
(This is closely related to the fact that we have employed
$r = 1$.). As in \cite{salm}, we choose 
$\gamma_1 =\sigma_3$, $\gamma_2 = \sigma_1$ 
($\sigma_i$ are Pauli matrices.)
and the projection operators $T_\mu^{(\epsilon)} = (1+\epsilon\gamma_\mu)/2$,
$\epsilon\in \{ -1,1\}$ then read $T_1^{(+)}={\rm diag}\{1,0\}$,
$T_1^{(-)}={\rm diag}\{0,1\}$, 
$T_2^{(\epsilon)}={1\over 2}{1\,\epsilon\choose\epsilon\,1}$.
It is now advantageous in order to simplify further results 
to define combinations of the original $\psi$ fields
in terms of which $T_2^{(\epsilon)}$ is diagonal 

\begin{eqnarray}
\label{B5}
\chi(x)&=&U \psi(x)\ =\ {1\over\sqrt{2}}
\left(\psi_1(x) + \psi_2(x),
\psi_1(x) - \psi_2(x)\right)^T\ \ \ ,
\\[0.3cm]
\bar\chi(x)&=&\bar\psi(x) U\ =\ {1\over\sqrt{2}}
\left(\bar\psi_1(x) + \bar\psi_2(x),
\bar\psi_1(x) - \bar\psi_2(x)\right)
\end{eqnarray}

with

\begin{eqnarray}
\label{B6}
U&=&{1\over\sqrt{2}}\left(
\begin{array}{cr}
1 & 1 \\ 1 & -1
\end{array}\right)
\ =\ U^{-1}\ \ \ .
\end{eqnarray}

\parindent1.5em

The discussion in this section proceeds via the following steps
indicated by subsections. In subsection 2.1, the Grassmann integration
is carried out exactly in two steps, first on the even and then
on the odd sublattices. In this process, a procedure based on 
graphical elements (vertices) is established for describing the 
results. Combinations of these graphical elements (vertex
clusters) stand in correspondence to contributions to the 
partition function (sum) $Z_\Lambda$. In subsection 2.2, 
rules for the calculation of the weight of the contribution
of an arbitrary vertex cluster are discussed. In subsection 
2.3, the fact that the weight of certain vertex clusters 
may vanish (due to certain 
cancellations among different terms contributing to the 
partition function) is exploited to interpret the graphical 
rules established in subsection 2.1 in terms of a loop model
picture. This finally allows to specify in subsection 2.4
the statistical model equivalent to the one-flavour lattice
Thirring model with Wilson fermions.\\

\subsection{The result of the Grassmann integration}

In the expression for the partition function (\ref{B4}) the 
integration on the even sublattice $\Lambda_e$ is now performed
first. As each lattice site $x$ supports 4 Grassmann variables 
terms which respect 

\parindent0.em

\begin{eqnarray}
\label{B7}
\sum_{l\ni x}\ k_l\ +\ 2 s_x\ + 4 t_x&=&4
\end{eqnarray}

can give a non-vanishing contribution only ($s_x = 0,1,2$ is the power of 
$M \bar\psi(x)\psi(x)$ and $t_x = 0,1$ is the power of 
$2 G \bar\psi_1(x) \bar\psi_2(x) \psi_1(x) \psi_2(x)$
in the expansion of the exponentials in eq.\ (\ref{B4}).). 
Below we give the results for all
local integrals related to a given lattice site $x\in\Lambda_e$ 
which allow non-vanishing contributions. We have relied on a 
purpose-written Mathematica program \cite{math} to do the 
necessary algebra. The results have been arranged in a way 
most suitable for the further discussion. The graphical rules 
below have to be interpreted as follows: the black
dot denotes any point $x$ on the even sublattice $\Lambda_e$ 
at which the Grassmann integration
is performed, a dashed line means $k_l = 0$ while thin and thick 
lines stand for $k_l = 1$ and $k_l = 2$ respectively.
A vertex is understood to be a lattice point with certain
admissible values of the occupation numbers $k_l$ on the 
attached links.
A given set of occupation numbers $k_l$ on the lattice $\Lambda$ uniquely
determines a certain diagram denoting a contribution 
to the partition function (As one can see from eq. (\ref{V1}),
we consider the cases $s_x = 2$, $t_x = 0$ and $s_x = 0$, $t_x = 1$
as unified into one.).
Finally, the first coordinate component of $x$ is understood to be the 
horizontal one while the second component runs vertically.\\

\parindent1.5em

There is one possible vertex with $s_x=2$ or $t_x=1$:\\

\parindent0.em
Vertex 1
\nopagebreak

\vspace{5mm}
\nopagebreak

\unitlength1.mm
\begin{picture}(150,27)
\put(0,12){
\unitlength1.mm
\begin{picture}(15,15)
\linethickness{0.15mm}
\put(12,7.5){\line(1,0){1}}
\put(10,7.5){\line(1,0){1}}
\put( 8,7.5){\line(1,0){1}}
\put( 6,7.5){\line(1,0){1}}
\put( 4,7.5){\line(1,0){1}}
\put( 2,7.5){\line(1,0){1}}
\put(7.5,12){\line(0,1){1}}
\put(7.5,10){\line(0,1){1}}
\put(7.5, 8){\line(0,1){1}}
\put(7.5, 6){\line(0,1){1}}
\put(7.5, 4){\line(0,1){1}}
\put(7.5, 2){\line(0,1){1}}
\put(7.5,7.5){\circle*{1.5}}
\end{picture}  }
\put(0,30){
\parbox[t]{15cm}{
\begin{eqnarray}
\label{V1}
&=&\ \ \ \int\prod_{\alpha=1}^2
d\psi_\alpha(x)\ d\bar\psi_\alpha(x)\ \Bigg\{ {1\over 2}\ 
\left(M \bar\psi(x)\psi(x)\right)^2
\ \ \ \ \ \ \ \ \ \ \ \ \ \ \ \ \ \ \ \ \nonumber\\[3mm]
&&\ \ \ \ \ +\ 2\ G\ \bar\psi_1(x)\bar\psi_2(x)
\psi_1(x)\psi_2(x)\Bigg\}\nonumber\\[3mm]
&=&\ \ \ M^2 - 2\ G 
\end{eqnarray} }}
\end{picture}

\vspace{2cm}

All further vertices have $s_x=0$ (vertices 2-20)
or $s_x=1$ (vertices 21-26) and $t_x=0$
(The results for the vertices 2-7 can also be found in
the paper by Salmhofer \cite{salm}, including the 
explicit steps of the derivation of the r.h.s.\ of 
eqs.\ (\ref{V3}) and (\ref{V5}).).\\

Vertex 2
\nopagebreak

\vspace{5mm}
\nopagebreak

\unitlength1.mm
\begin{picture}(150,20)
\put(0,5){
\unitlength1.mm
\begin{picture}(15,15)
\linethickness{0.15mm}
\put(12,7.5){\line(1,0){1}}
\put(10,7.5){\line(1,0){1}}
\put( 8,7.5){\line(1,0){1}}
\put( 6,7.5){\line(1,0){1}}
\put( 4,7.5){\line(1,0){1}}
\put( 2,7.5){\line(1,0){1}}
\linethickness{0.6mm}
\put(7.5,2){\line(0,1){11}}
\put(7.5,7.5){\circle*{1.5}}
\end{picture} }
\put(0,21){
\parbox[t]{15cm}{
\begin{eqnarray}
\label{V2}
&=&\ \ \ \ \ 
\ \chi_1(x-e_2)\ \bar\chi_1(x+e_2)
\ \chi_2(x+e_2)\ \bar\chi_2(x-e_2)
\ \ \ \ \ \ \ \ \ \ \ \ \ \
\end{eqnarray}  }}
\end{picture}

Vertex 3
\nopagebreak

\vspace{5mm}
\nopagebreak

\unitlength1.mm
\begin{picture}(150,20)
\put(0,5){
\unitlength1.mm
\begin{picture}(15,15)
\linethickness{0.15mm}
\put(7.5,12){\line(0,1){1}}
\put(7.5,10){\line(0,1){1}}
\put(7.5, 8){\line(0,1){1}}
\put(7.5, 6){\line(0,1){1}}
\put(7.5, 4){\line(0,1){1}}
\put(7.5, 2){\line(0,1){1}}
\linethickness{0.6mm}
\put(2,7.5){\line(1,0){11}}
\put(7.5,7.5){\circle*{1.5}}
\end{picture} }
\put(0,21){
\parbox[t]{15cm}{
\begin{eqnarray}
\label{V3}
&=&\ \ \ \ \ 
\ \psi_1(x-e_1)\ \bar\psi_1(x+e_1)
\ \psi_2(x+e_1)\ \bar\psi_2(x-e_1)
\ \ \ \ \ \ \ \ \ \ \ \ \ \
\end{eqnarray}  }}
\end{picture}

Vertex 4
\nopagebreak

\vspace{5mm}
\nopagebreak

\unitlength1.mm
\begin{picture}(150,20)
\put(0,5){
\unitlength1.mm
\begin{picture}(15,15)
\linethickness{0.15mm}
\put( 6,7.5){\line(1,0){1}}
\put( 4,7.5){\line(1,0){1}}
\put( 2,7.5){\line(1,0){1}}
\put(7.5,12){\line(0,1){1}}
\put(7.5,10){\line(0,1){1}}
\put(7.5, 8){\line(0,1){1}}
\linethickness{0.6mm}
\put(7.5,7.5){\line(1,0){5.5}}
\put(7.5,2){\line(0,1){5.5}}
\put(7.5,7.5){\circle*{1.5}}
\end{picture} }
\put(0,21){
\parbox[t]{15cm}{
\begin{eqnarray}
\label{V4}
&=&\ \ \ -\ \psi_2(x+e_1)\ \bar\chi_2(x-e_2)
\ \chi_1(x-e_2)\ \bar\psi_1(x+e_1)/2
\ \ \ \ \ \ \ \ \ \ \
\end{eqnarray}  }}
\end{picture}

Vertex 5
\nopagebreak

\vspace{5mm}
\nopagebreak

\unitlength1.mm
\begin{picture}(150,20)
\put(0,5){
\unitlength1.mm
\begin{picture}(15,15)
\linethickness{0.15mm}
\put(12,7.5){\line(1,0){1}}
\put(10,7.5){\line(1,0){1}}
\put( 8,7.5){\line(1,0){1}}
\put(7.5, 6){\line(0,1){1}}
\put(7.5, 4){\line(0,1){1}}
\put(7.5, 2){\line(0,1){1}}
\linethickness{0.6mm}
\put(2,7.5){\line(1,0){5.5}}
\put(7.5,7.5){\line(0,1){5.5}}
\put(7.5,7.5){\circle*{1.5}}
\end{picture} }
\put(0,21){
\parbox[t]{15cm}{
\begin{eqnarray}
\label{V5}
&=&\ \ \ -\ \psi_1(x-e_1)\ \bar\chi_1(x+e_2)
\ \chi_2(x+e_2)\ \bar\psi_2(x-e_1)/2
\ \ \ \ \ \ \ \ \ \ \
\end{eqnarray}  }}
\end{picture}

Vertex 6
\nopagebreak

\vspace{5mm}
\nopagebreak

\unitlength1.mm
\begin{picture}(150,20)
\put(0,5){
\unitlength1.mm
\begin{picture}(15,15)
\linethickness{0.15mm}
\put( 6,7.5){\line(1,0){1}}
\put( 4,7.5){\line(1,0){1}}
\put( 2,7.5){\line(1,0){1}}
\put(7.5, 6){\line(0,1){1}}
\put(7.5, 4){\line(0,1){1}}
\put(7.5, 2){\line(0,1){1}}
\linethickness{0.6mm}
\put(7.5,7.5){\line(1,0){5.5}}
\put(7.5,7.5){\line(0,1){5.5}}
\put(7.5,7.5){\circle*{1.5}}
\end{picture} }
\put(0,21){
\parbox[t]{15cm}{
\begin{eqnarray}
\label{V6}
&=&\ \ \ \ 
\ \psi_2(x+e_1)\ \bar\chi_1(x+e_2)
\ \chi_2(x+e_2)\ \bar\psi_1(x+e_1)/2
\ \ \ \ \ \ \ \ \ \ \ \
\end{eqnarray}  }}
\end{picture}

Vertex 7
\nopagebreak

\vspace{5mm}
\nopagebreak

\unitlength1.mm
\begin{picture}(150,20)
\put(0,5){
\unitlength1.mm
\begin{picture}(15,15)
\linethickness{0.15mm}
\put(12,7.5){\line(1,0){1}}
\put(10,7.5){\line(1,0){1}}
\put( 8,7.5){\line(1,0){1}}
\put(7.5,12){\line(0,1){1}}
\put(7.5,10){\line(0,1){1}}
\put(7.5, 8){\line(0,1){1}}
\linethickness{0.6mm}
\put(2,7.5){\line(1,0){5.5}}
\put(7.5,2){\line(0,1){5.5}}
\put(7.5,7.5){\circle*{1.5}}
\end{picture} }
\put(0,21){
\parbox[t]{15cm}{
\begin{eqnarray}
\label{V7}
&=&\ \ \ \ 
\ \psi_1(x-e_1)\ \bar\chi_2(x-e_2)
\ \chi_1(x-e_2)\ \bar\psi_2(x-e_1)/2
\ \ \ \ \ \ \ \ \ \ \ \
\end{eqnarray}  }}
\end{picture}

Vertex 8
\nopagebreak

\vspace{5mm}
\nopagebreak

\unitlength1.mm
\begin{picture}(150,25)
\put(0,10){
\unitlength1.mm
\begin{picture}(15,15)
\linethickness{0.15mm}
\put(12,7.5){\line(1,0){1}}
\put(10,7.5){\line(1,0){1}}
\put( 8,7.5){\line(1,0){1}}
\put(7.5,2){\line(0,1){11}}
\linethickness{0.6mm}
\put(2,7.5){\line(1,0){5.5}}
\put(7.5,7.5){\circle*{1.5}}
\end{picture} }
\put(0,26){
\parbox[t]{15cm}{
\begin{eqnarray}
\label{V8}
&=&\ \ \ \ [\psi_1(x-e_1)\ \bar\chi_1(x+e_2)
\ \chi_1(x-e_2)\ \bar\psi_2(x-e_1)\ \ \ \ \ \ \ \ \ \ \ \ \ \ \ \
\nonumber\\[0.3cm]
&&\ \ \ \ -\ \psi_1(x-e_1)\ \bar\chi_2(x-e_2)
\ \chi_2(x+e_2)\ \bar\psi_2(x-e_1)]/2
\end{eqnarray}  }}
\end{picture}

Vertex 9
\nopagebreak

\vspace{5mm}
\nopagebreak

\unitlength1.mm
\begin{picture}(150,25)
\put(0,10){
\unitlength1.mm
\begin{picture}(15,15)
\linethickness{0.15mm}
\put( 6,7.5){\line(1,0){1}}
\put( 4,7.5){\line(1,0){1}}
\put( 2,7.5){\line(1,0){1}}
\put(7.5,2){\line(0,1){11}}
\linethickness{0.6mm}
\put(7.5,7.5){\line(1,0){5.5}}
\put(7.5,7.5){\circle*{1.5}}
\end{picture} }
\put(0,26){
\parbox[t]{15cm}{
\begin{eqnarray}
\label{V9}
&=&\ \ \ \ [\psi_2(x+e_1)\ \bar\chi_1(x+e_2)
\ \chi_1(x-e_2)\ \bar\psi_1(x+e_1)\ \ \ \ \ \ \ \ \ \ \ \ \ \ \ \ 
\nonumber\\[0.3cm]
&&\ \ \ \ -\ \psi_2(x+e_1)\ \bar\chi_2(x-e_2)
\ \chi_2(x+e_2)\ \bar\psi_1(x+e_1)]/2
\end{eqnarray}  }}
\end{picture}

Vertex 10
\nopagebreak

\vspace{5mm}
\nopagebreak

\unitlength1.mm
\begin{picture}(150,25)
\put(0,10){
\unitlength1.mm
\begin{picture}(15,15)
\linethickness{0.15mm}
\put(2,7.5){\line(1,0){11}}
\put(7.5, 6){\line(0,1){1}}
\put(7.5, 4){\line(0,1){1}}
\put(7.5, 2){\line(0,1){1}}
\linethickness{0.6mm}
\put(7.5,7.5){\line(0,1){5.5}}
\put(7.5,7.5){\circle*{1.5}}
\end{picture} }
\put(0,26){
\parbox[t]{15cm}{
\begin{eqnarray}
\label{V10}
&=&\ \ \ \ [\chi_2(x+e_2)\ \bar\psi_1(x+e_1)
\ \psi_1(x-e_1)\ \bar\chi_1(x+e_2)\ \ \ \ \ \ \ \ \ \ \ \ \ \ \ \ 
\nonumber\\[0.3cm]
&&\ \ \ \ -\ \chi_2(x+e_2)\ \bar\psi_2(x-e_1)
\ \psi_2(x+e_1)\ \bar\chi_1(x+e_2)]/2
\end{eqnarray}  }}
\end{picture}

Vertex 11
\nopagebreak

\vspace{5mm}
\nopagebreak

\unitlength1.mm
\begin{picture}(150,25)
\put(0,10){
\unitlength1.mm
\begin{picture}(15,15)
\linethickness{0.15mm}
\put(2,7.5){\line(1,0){11}}
\put(7.5,12){\line(0,1){1}}
\put(7.5,10){\line(0,1){1}}
\put(7.5, 8){\line(0,1){1}}
\linethickness{0.6mm}
\put(7.5,2){\line(0,1){5.5}}
\put(7.5,7.5){\circle*{1.5}}
\end{picture} }
\put(0,26){
\parbox[t]{15cm}{
\begin{eqnarray}
\label{V11}
&=&\ \ \ \ [\chi_1(x-e_2)\ \bar\psi_1(x+e_1)
\ \psi_1(x-e_1)\ \bar\chi_2(x-e_2)\ \ \ \ \ \ \ \ \ \ \ \ \ \ \ \ 
\nonumber\\[0.3cm]
&&\ \ \ \ -\ \chi_1(x-e_2)\ \bar\psi_2(x-e_1)
\ \psi_2(x+e_1)\ \bar\chi_2(x-e_2)]/2
\end{eqnarray}  }}
\end{picture}

Vertex 12
\nopagebreak

\vspace{5mm}
\nopagebreak

\unitlength1.mm
\begin{picture}(150,25)
\put(0,10){
\unitlength1.mm
\begin{picture}(15,15)
\linethickness{0.15mm}
\put(7.5,12){\line(0,1){1}}
\put(7.5,10){\line(0,1){1}}
\put(7.5, 8){\line(0,1){1}}
\put(7.5,7.5){\line(1,0){5.5}}
\put(7.5,2){\line(0,1){5.5}}
\linethickness{0.6mm}
\put(2,7.5){\line(1,0){5.5}}
\put(7.5,7.5){\circle*{1.5}}
\end{picture} }
\put(0,26){
\parbox[t]{15cm}{
\begin{eqnarray}
\label{V12}
&=&\ \ \ \ [\psi_1(x-e_1)\ \bar\psi_1(x+e_1)
\ \chi_1(x-e_2)\ \bar\psi_2(x-e_1)\ \ \ \ \ \ \ \ \ \ \ \ \ \ \ 
\nonumber\\[0.3cm]
&&\ \ \ \ +\ \psi_1(x-e_1)\ \bar\chi_2(x-e_2)
\ \psi_2(x+e_1)\ \bar\psi_2(x-e_1)]/\sqrt{2}
\end{eqnarray}  }}
\end{picture}

Vertex 13
\nopagebreak

\vspace{5mm}
\nopagebreak

\unitlength1.mm
\begin{picture}(150,25)
\put(0,10){
\unitlength1.mm
\begin{picture}(15,15)
\linethickness{0.15mm}
\put( 6,7.5){\line(1,0){1}}
\put( 4,7.5){\line(1,0){1}}
\put( 2,7.5){\line(1,0){1}}
\put(7.5,2){\line(0,1){5.5}}
\put(7.5,7.5){\line(1,0){5.5}}
\linethickness{0.6mm}
\put(7.5,7.5){\line(0,1){5.5}}
\put(7.5,7.5){\circle*{1.5}}
\end{picture} }
\put(0,26){
\parbox[t]{15cm}{
\begin{eqnarray}
\label{V13}
&=&\ \ \ \ [\chi_2(x+e_2)\ \bar\psi_1(x+e_1)
\ \chi_1(x-e_2)\ \bar\chi_1(x+e_2)\ \ \ \ \ \ \ \ \ \ \ \ \ \ \ 
\nonumber\\[0.3cm]
&&\ \ \ \ +\ \chi_2(x+e_2)\ \bar\chi_2(x-e_2)
\ \psi_2(x+e_1)\ \bar\chi_1(x+e_2)]/\sqrt{2}
\end{eqnarray}  }}
\end{picture}

Vertex 14
\nopagebreak

\vspace{5mm}
\nopagebreak

\unitlength1.mm
\begin{picture}(150,25)
\put(0,10){
\unitlength1.mm
\begin{picture}(15,15)
\linethickness{0.15mm}
\put(7.5, 6){\line(0,1){1}}
\put(7.5, 4){\line(0,1){1}}
\put(7.5, 2){\line(0,1){1}}
\put(2,7.5){\line(1,0){5.5}}
\put(7.5,7.5){\line(0,1){5.5}}
\linethickness{0.6mm}
\put(7.5,7.5){\line(1,0){5.5}}
\put(7.5,7.5){\circle*{1.5}}
\end{picture} }
\put(0,26){
\parbox[t]{15cm}{
\begin{eqnarray}
\label{V14}
&=&\ \ \ \ [\psi_2(x+e_1)\ \bar\psi_2(x-e_1)
\ \chi_2(x+e_2)\ \bar\psi_1(x+e_1)\ \ \ \ \ \ \ \ \ \ \ \ \ \ \ 
\nonumber\\[0.3cm]
&&\ \ \ \ +\ \psi_2(x+e_1)\ \bar\chi_1(x+e_2)
\ \psi_1(x-e_1)\ \bar\psi_1(x+e_1)]/\sqrt{2}
\end{eqnarray}  }}
\end{picture}

Vertex 15
\nopagebreak

\vspace{5mm}
\nopagebreak

\unitlength1.mm
\begin{picture}(150,25)
\put(0,10){
\unitlength1.mm
\begin{picture}(15,15)
\linethickness{0.15mm}
\put(12,7.5){\line(1,0){1}}
\put(10,7.5){\line(1,0){1}}
\put( 8,7.5){\line(1,0){1}}
\put(7.5,7.5){\line(0,1){5.5}}
\put(2,7.5){\line(1,0){5.5}}
\linethickness{0.6mm}
\put(7.5,2){\line(0,1){5.5}}
\put(7.5,7.5){\circle*{1.5}}
\end{picture} }
\put(0,26){
\parbox[t]{15cm}{
\begin{eqnarray}
\label{V15}
&=&\ \ \ [\chi_1(x-e_2)\ \bar\psi_2(x-e_1)
\ \chi_2(x+e_2)\ \bar\chi_2(x-e_2)\ \ \ \ \ \ \ \ \ \ \ \ \ \ \ \ 
\nonumber\\[0.3cm]
&&\ \ \ \ +\ \chi_1(x-e_2)\ \bar\chi_1(x+e_2)
\ \psi_1(x-e_1)\ \bar\chi_2(x-e_2)]/\sqrt{2}
\end{eqnarray}  }}
\end{picture}

Vertex 16
\nopagebreak

\vspace{5mm}
\nopagebreak

\unitlength1.mm
\begin{picture}(150,25)
\put(0,10){
\unitlength1.mm
\begin{picture}(15,15)
\linethickness{0.15mm}
\put(7.5, 6){\line(0,1){1}}
\put(7.5, 4){\line(0,1){1}}
\put(7.5, 2){\line(0,1){1}}
\put(7.5,7.5){\line(1,0){5.5}}
\put(7.5,7.5){\line(0,1){5.5}}
\linethickness{0.6mm}
\put(2,7.5){\line(1,0){5.5}}
\put(7.5,7.5){\circle*{1.5}}
\end{picture} }
\put(0,26){
\parbox[t]{15cm}{
\begin{eqnarray}
\label{V16}
&=&\ \ \ \ [-\ \psi_1(x-e_1)\ \bar\psi_1(x+e_1)
\ \chi_2(x+e_2)\ \bar\psi_2(x-e_1)\ \ \ \ \ \ \ \ \ \ \
\nonumber\\[0.3cm]
&&\ \ \ \ +\ \psi_1(x-e_1)\ \bar\chi_1(x+e_2)
\ \psi_2(x+e_1)\ \bar\psi_2(x-e_1)]/\sqrt{2}
\end{eqnarray}  }}
\end{picture}

Vertex 17
\nopagebreak

\vspace{5mm}
\nopagebreak

\unitlength1.mm
\begin{picture}(150,25)
\put(0,10){
\unitlength1.mm
\begin{picture}(15,15)
\linethickness{0.15mm}
\put( 6,7.5){\line(1,0){1}}
\put( 4,7.5){\line(1,0){1}}
\put( 2,7.5){\line(1,0){1}}
\put(7.5,7.5){\line(0,1){5.5}}
\put(7.5,7.5){\line(1,0){5.5}}
\linethickness{0.6mm}
\put(7.5,2){\line(0,1){5.5}}
\put(7.5,7.5){\circle*{1.5}}
\end{picture} }
\put(0,26){
\parbox[t]{15cm}{
\begin{eqnarray}
\label{V17}
&=&\ \ \ \ [\chi_1(x-e_2)\ \bar\psi_1(x+e_1)
\ \chi_2(x+e_2)\ \bar\chi_2(x-e_2)\ \ \ \ \ \ \ \ \ \ \ \ \ \ \ \
\nonumber\\[0.3cm]
&&\ \ \ \ -\ \chi_1(x-e_2)\ \bar\chi_1(x+e_2)
\ \psi_2(x+e_1)\ \bar\chi_2(x-e_2)]/\sqrt{2}
\end{eqnarray}  }}
\end{picture}

Vertex 18
\nopagebreak

\vspace{5mm}
\nopagebreak

\unitlength1.mm
\begin{picture}(150,25)
\put(0,10){
\unitlength1.mm
\begin{picture}(15,15)
\linethickness{0.15mm}
\put(12,7.5){\line(1,0){1}}
\put(10,7.5){\line(1,0){1}}
\put( 8,7.5){\line(1,0){1}}
\put(7.5,2){\line(0,1){5.5}}
\put(2,7.5){\line(1,0){5.5}}
\linethickness{0.6mm}
\put(7.5,7.5){\line(0,1){5.5}}
\put(7.5,7.5){\circle*{1.5}}
\end{picture} }
\put(0,26){
\parbox[t]{15cm}{
\begin{eqnarray}
\label{V18}
&=&\ \ \ \ [-\ \chi_2(x+e_2)\ \bar\psi_2(x-e_1)
\ \chi_1(x-e_2)\ \bar\chi_1(x+e_2)\ \ \ \ \ \ \ \ \ \ \ \ 
\nonumber\\[0.3cm]
&&\ \ \ \ +\ \chi_2(x+e_2)\ \bar\chi_2(x-e_2)
\ \psi_1(x-e_1)\ \bar\chi_1(x+e_2)]/\sqrt{2}
\end{eqnarray}  }}
\end{picture}

Vertex 19
\nopagebreak

\vspace{5mm}
\nopagebreak

\unitlength1.mm
\begin{picture}(150,25)
\put(0,10){
\unitlength1.mm
\begin{picture}(15,15)
\linethickness{0.15mm}
\put(7.5,12){\line(0,1){1}}
\put(7.5,10){\line(0,1){1}}
\put(7.5, 8){\line(0,1){1}}
\put(2,7.5){\line(1,0){5.5}}
\put(7.5,2){\line(0,1){5.5}}
\linethickness{0.6mm}
\put(7.5,7.5){\line(1,0){5.5}}
\put(7.5,7.5){\circle*{1.5}}
\end{picture} }
\put(0,26){
\parbox[t]{15cm}{
\begin{eqnarray}
\label{V19}
&=&\ \ \ \ [\psi_2(x+e_1)\ \bar\psi_2(x-e_1)
\ \chi_1(x-e_2)\ \bar\psi_1(x+e_1)\ \ \ \ \ \ \ \ \ \ \ \ \ \ \ 
\nonumber\\[0.3cm]
&&\ \ \ \ -\ \psi_2(x+e_1)\ \bar\chi_2(x-e_2)
\ \psi_1(x-e_1)\ \bar\psi_1(x+e_1)]/\sqrt{2}
\end{eqnarray}  }}
\end{picture}

Vertex 20
\nopagebreak

\vspace{5mm}
\nopagebreak

\unitlength1.mm
\begin{picture}(150,65)
\put(0,50){
\unitlength1.mm
\begin{picture}(15,15)
\linethickness{0.15mm}
\put(2,7.5){\line(1,0){11}}
\put(7.5,2){\line(0,1){11}}
\put(7.5,7.5){\circle*{1.5}}
\end{picture} }
\put(0,66){
\parbox[t]{15cm}{
\begin{eqnarray}
\label{V20}
&=&-\ [\ \ \ \ \chi_1(x-e_2)\ \bar\psi_1(x+e_1)\ \chi_2(x+e_2)
\ \bar\psi_2(x-e_1)
\ \ \ \ \ \ \ \ \ \ \ \ \ \ \ \ \ \nonumber\\[0.3cm]
&&\ \ \ \ \ +\ \psi_1(x-e_1)\ \bar\chi_1(x+e_2)
\ \chi_1(x-e_2)\ \bar\psi_1(x+e_1)/2\nonumber\\[0.3cm]
&&\ \ \ \ \ +\ \psi_1(x-e_1)\ \bar\chi_2(x-e_2)
\ \chi_2(x+e_2)\ \bar\psi_1(x+e_1)/2\nonumber\\[0.3cm]
&&\ \ \ \ \ +\ \psi_2(x+e_1)\ \bar\chi_1(x+e_2)
\ \chi_1(x+e_2)\ \bar\psi_2(x-e_1)/2\nonumber\\[0.3cm]
&&\ \ \ \ \ +\ \psi_2(x+e_1)\ \bar\chi_2(x-e_2)
\ \chi_2(x+e_2)\ \bar\psi_2(x-e_1)/2\nonumber\\[0.3cm]
&&\ \ \ \ \ +\ \psi_1(x-e_1)\ \bar\chi_1(x+e_2)\ \psi_2(x+e_1)
\ \bar\chi_2(x-e_2) ]
\end{eqnarray}  }}
\end{picture}

Vertex 21
\nopagebreak

\vspace{5mm}
\nopagebreak

\unitlength1.mm
\begin{picture}(150,20)
\put(0,5){
\unitlength1.mm
\begin{picture}(15,15)
\linethickness{0.15mm}
\put(12,7.5){\line(1,0){1}}
\put(10,7.5){\line(1,0){1}}
\put( 8,7.5){\line(1,0){1}}
\put( 6,7.5){\line(1,0){1}}
\put( 4,7.5){\line(1,0){1}}
\put( 2,7.5){\line(1,0){1}}
\put(7.5,2){\line(0,1){11}}
\put(7.5,7.5){\circle*{1.5}}
\end{picture} }
\put(0,21){
\parbox[t]{15cm}{
\begin{eqnarray}
\label{V21}
&=&M\ [\chi_1(x-e_2)\ \bar\chi_1(x+e_2)\ +
\ \chi_2(x+e_2)\ \bar\chi_2(x-e_2) ]\ \ \ \ \ \ \ 
\end{eqnarray}  }}
\end{picture}

Vertex 22
\nopagebreak

\vspace{5mm}
\nopagebreak

\unitlength1.mm
\begin{picture}(150,20)
\put(0,5){
\unitlength1.mm
\begin{picture}(15,15)
\linethickness{0.15mm}
\put(7.5,12){\line(0,1){1}}
\put(7.5,10){\line(0,1){1}}
\put(7.5, 8){\line(0,1){1}}
\put(7.5, 6){\line(0,1){1}}
\put(7.5, 4){\line(0,1){1}}
\put(7.5, 2){\line(0,1){1}}
\put(2,7.5){\line(1,0){11}}
\put(7.5,7.5){\circle*{1.5}}
\end{picture} }
\put(0,21){
\parbox[t]{15cm}{
\begin{eqnarray}
\label{V22}
&=&M\  [\psi_1(x-e_1)\ \bar\psi_1(x+e_1)\ +
\ \psi_2(x+e_1)\ \bar\psi_2(x-e_1) ]\ \ \ \ \ \ \ 
\end{eqnarray}  }}
\end{picture}

Vertex 23
\nopagebreak

\vspace{5mm}
\nopagebreak

\unitlength1.mm
\begin{picture}(150,20)
\put(0,5){
\unitlength1.mm
\begin{picture}(15,15)
\linethickness{0.15mm}
\put( 6,7.5){\line(1,0){1}}
\put( 4,7.5){\line(1,0){1}}
\put( 2,7.5){\line(1,0){1}}
\put(7.5,12){\line(0,1){1}}
\put(7.5,10){\line(0,1){1}}
\put(7.5, 8){\line(0,1){1}}
\put(7.5,7.5){\line(1,0){5.5}}
\put(7.5,2){\line(0,1){5.5}}
\put(7.5,7.5){\circle*{1.5}}
\end{picture} }
\put(0,21){
\parbox[t]{15cm}{
\begin{eqnarray}
\label{V23}
&=&M\ [\chi_1(x-e_2)\ \bar\psi_1(x+e_1)\ -
\ \psi_2(x+e_1)\ \bar\chi_2(x-e_2) ]/\sqrt{2}\ \ 
\end{eqnarray}  }}
\end{picture}

Vertex 24
\nopagebreak

\vspace{5mm}
\nopagebreak

\unitlength1.mm
\begin{picture}(150,20)
\put(0,5){
\unitlength1.mm
\begin{picture}(15,15)
\linethickness{0.15mm}
\put(12,7.5){\line(1,0){1}}
\put(10,7.5){\line(1,0){1}}
\put( 8,7.5){\line(1,0){1}}
\put(7.5, 6){\line(0,1){1}}
\put(7.5, 4){\line(0,1){1}}
\put(7.5, 2){\line(0,1){1}}
\put(2,7.5){\line(1,0){5.5}}
\put(7.5,7.5){\line(0,1){5.5}}
\put(7.5,7.5){\circle*{1.5}}
\end{picture} }
\put(0,21){
\parbox[t]{15cm}{
\begin{eqnarray}
\label{V24}
&=&M\ [ \psi_1(x-e_1)\ \bar\chi_1(x+e_2)\ -
\ \chi_2(x+e_2)\ \bar\psi_2(x-e_1)]/\sqrt{2}\ \ 
\end{eqnarray}  }}
\end{picture}

Vertex 25
\nopagebreak

\vspace{5mm}
\nopagebreak

\unitlength1.mm
\begin{picture}(150,20)
\put(0,5){
\unitlength1.mm
\begin{picture}(15,15)
\linethickness{0.15mm}
\put( 6,7.5){\line(1,0){1}}
\put( 4,7.5){\line(1,0){1}}
\put( 2,7.5){\line(1,0){1}}
\put(7.5, 6){\line(0,1){1}}
\put(7.5, 4){\line(0,1){1}}
\put(7.5, 2){\line(0,1){1}}
\put(7.5,7.5){\line(1,0){5.5}}
\put(7.5,7.5){\line(0,1){5.5}}
\put(7.5,7.5){\circle*{1.5}}
\end{picture} }
\put(0,21){
\parbox[t]{15cm}{
\begin{eqnarray}
\label{V25}
&=&M\ [\chi_2(x+e_2)\ \bar\psi_1(x+e_1)\ +
\ \psi_2(x+e_1)\ \bar\chi_1(x+e_2) ]/\sqrt{2}\ \
\end{eqnarray}  }}
\end{picture}

Vertex 26
\nopagebreak

\vspace{5mm}
\nopagebreak

\unitlength1.mm
\begin{picture}(150,20)
\put(0,5){
\unitlength1.mm
\begin{picture}(15,15)
\linethickness{0.15mm}
\put(12,7.5){\line(1,0){1}}
\put(10,7.5){\line(1,0){1}}
\put( 8,7.5){\line(1,0){1}}
\put(7.5,12){\line(0,1){1}}
\put(7.5,10){\line(0,1){1}}
\put(7.5, 8){\line(0,1){1}}
\put(2,7.5){\line(1,0){5.5}}
\put(7.5,2){\line(0,1){5.5}}
\put(7.5,7.5){\circle*{1.5}}
\end{picture} }
\put(0,21){
\parbox[t]{15cm}{
\begin{eqnarray}
\label{V26}
&=&M\ [ \chi_1(x-e_2)\ \bar\psi_2(x-e_1)\ +
\ \psi_1(x-e_1)\ \bar\chi_2(x-e_2) ]/\sqrt{2}\
\end{eqnarray}  }}
\end{picture}

In order to achieve further understanding it is 
useful to assign the remaining 
$\bar\psi$, $\psi$ fields on the odd sublattice $\Lambda_o$ certain  
graphical symbols as follows (The full black dot denotes a 
point $x$ on the even sublattice while the hollow circle stands 
for the point on the odd sublattice which the argument
of the fields relates to.).\\

\vspace{3mm}

\parindent0.em

\unitlength1.mm
\begin{picture}(150,7)
\put(0,5){
\unitlength1.mm
\begin{picture}(8.25,2)
\linethickness{0.15mm}
\put(2,1){\vector(1,0){3.75}}
\put(7.5,1){\line(-1,0){1.75}}
\put(1,1){\circle{2}}
\put(7.5,1){\circle*{1.5}}
\end{picture}  }
\put(70,5){
\unitlength1.mm
\begin{picture}(8.25,2)
\linethickness{0.15mm}
\put(7.5,1){\vector(-1,0){3.75}}
\put(2,1){\line(1,0){1.75}}
\put(1,1){\circle{2}}
\put(7.5,1){\circle*{1.5}}
\end{picture}   }
\put(0,6){
\parbox[t]{15cm}{
\parbox{6cm}{
\begin{eqnarray}
\label{F1}
=&\bar\psi_2(x-e_1)&
\end{eqnarray}  }
\parbox{7mm}{\ }
\parbox{6cm}{
\begin{eqnarray}
\label{F2}
=&\psi_1(x-e_1)&
\end{eqnarray}  }  }}
\end{picture} 

\vspace{6mm}

\unitlength1.mm
\begin{picture}(150,7)
\put(0,5){
\unitlength1.mm
\begin{picture}(8.25,2)
\linethickness{0.15mm}
\put(0.75,1){\vector(1,0){3.75}}
\put(6.25,1){\line(-1,0){1.75}}
\put(7.25,1){\circle{2}}
\put(0.75,1){\circle*{1.5}}
\end{picture}   }
\put(70,5){
\unitlength1.mm
\begin{picture}(8.25,2)
\linethickness{0.15mm}
\put(6.25,1){\vector(-1,0){3.75}}
\put(0.75,1){\line(1,0){1.75}}
\put(7.25,1){\circle{2}}
\put(0.75,1){\circle*{1.5}}
\end{picture}  }
\put(0,6){
\parbox[t]{15cm}{
\parbox{6cm}{
\begin{eqnarray}
\label{F3}
=&\psi_2(x+e_1)&
\end{eqnarray}  }
\parbox{7mm}{\ }
\parbox{6cm}{
\begin{eqnarray}
\label{F4}
=&\bar\psi_1(x+e_1)&
\end{eqnarray}  } }}
\end{picture}

\vspace{3mm}

\unitlength1.mm
\begin{picture}(150,8)
\put(3,0){
\unitlength1.mm
\begin{picture}(2,8.25)
\linethickness{0.15mm}
\put(1,2){\vector(0,1){3.75}}
\put(1,7.5){\line(0,-1){1.75}}
\put(1,1){\circle{2}}
\put(1,7.5){\circle*{1.5}}
\end{picture}  }
\put(73,0){
\unitlength1.mm
\begin{picture}(2,8.25)
\linethickness{0.15mm}
\put(1,7.5){\vector(0,-1){3.75}}
\put(1,2){\line(0,1){1.75}}
\put(1,1){\circle{2}}
\put(1,7.5){\circle*{1.5}}
\end{picture}    }
\put(0,4){
\parbox[t]{15cm}{
\parbox{6cm}{
\begin{eqnarray}
\label{F5}
=&\bar\chi_2(x-e_2)&
\end{eqnarray}  }
\parbox{7mm}{\ }
\parbox{6cm}{
\begin{eqnarray}
\label{F6}
=&\chi_1(x-e_2)&
\end{eqnarray}  } }}
\end{picture} 

\vspace{4mm}

\unitlength1.mm
\begin{picture}(150,8)
\put(3,0){
\unitlength1.mm
\begin{picture}(2,8.25)
\linethickness{0.15mm}
\put(1,0.75){\vector(0,1){3.75}}
\put(1,6.25){\line(0,-1){1.75}}
\put(1,7.25){\circle{2}}
\put(1,0.75){\circle*{1.5}}
\end{picture}  }
\put(73,0){
\unitlength1.mm
\begin{picture}(2,8.25)
\linethickness{0.15mm}
\put(1,6.25){\vector(0,-1){3.75}}
\put(1,0.75){\line(0,1){1.75}}
\put(1,7.25){\circle{2}}
\put(1,0.75){\circle*{1.5}}
\end{picture}    }
\put(0,4){
\parbox[t]{15cm}{
\parbox{6cm}{
\begin{eqnarray}
\label{F7}
=&\chi_2(x+e_2)&
\end{eqnarray}  }
\parbox{7mm}{\ }
\parbox{6cm}{
\begin{eqnarray}
\label{F8}
=&\bar\chi_1(x+e_2)&
\end{eqnarray}  }  }}
\end{picture} 

\vspace{7mm}

As a rule, an arrow flowing out of a point of the odd sublattice
(hollow circles) denotes a $\bar\psi$, $\bar\chi$ field variable while 
an arrow flowing into a point on the odd sublattice relates 
to a $\psi$, $\chi$ field variable. From this it is already
clear that only those products of (four) Grassmann fields 
at any point of the odd sublattice allow non-vanishing 
contributions to the partition function $Z_\Lambda$ 
which have two incoming and two outgoing arrows in their
graphical symbols. We will refer to this fact as the 
vertex arrow rule.
The symbols for the fields on the odd sublattice can now be 
used in a natural way to construct further graphical building blocks.
As will become clear further below we choose signs in accordance 
with eqs.\ (\ref{V21})-(\ref{V26}) (for $M = 1$).\\[5mm]

\unitlength1.mm
\begin{picture}(150,30)
\put(6.5,15){
\unitlength1.mm
\begin{picture}(2,15)
\linethickness{0.15mm}
\put(1,2){\line(0,1){11}}
\put(1,14){\circle{2}}
\put(1,7.5){\circle*{1.5}}
\put(1,1){\circle{2}}
\end{picture}  }
\put(51.5,15){
\unitlength1.mm
\begin{picture}(2,15)
\linethickness{0.15mm}
\put(1,2){\vector(0,1){3.75}}
\put(1,7.5){\line(0,-1){1.75}}
\put(1,7.5){\vector(0,1){3.75}}
\put(1,13){\line(0,-1){1.75}}
\put(1,1){\circle{2}}
\put(1,7.5){\circle*{1.5}}
\put(1,14){\circle{2}}
\end{picture}  }
\put(106.5,15){
\unitlength1.mm
\begin{picture}(2,15)
\linethickness{0.15mm}
\put(1,13){\vector(0,-1){3.75}}
\put(1,7.5){\line(0,1){1.75}}
\put(1,7.5){\vector(0,-1){3.75}}
\put(1,2){\line(0,1){1.75}}
\put(1,14){\circle{2}}
\put(1,7.5){\circle*{1.5}}
\put(1,1){\circle{2}}
\end{picture}  }
\put(0,30.75){
\parbox[t]{15cm}{
\begin{eqnarray}
\label{S1}
=&&\ \ \ \; \hspace{4.65cm}+\nonumber\\[11.5mm]
=&&\ \ \ \ \chi_2(x+e_2)\ \bar\chi_2(x-e_2)\ \ \ \ \ +\ \ \ \ 
\ \chi_1(x-e_2)\ \bar\chi_1(x+e_2)\ \ \
\end{eqnarray} }}
\end{picture} 

\vspace{7mm}

\unitlength1.mm
\begin{picture}(150,20)
\put(0,18){
\unitlength1.mm
\begin{picture}(15,2)
\linethickness{0.15mm}
\put(2,1){\line(1,0){11}}
\put(1,1){\circle{2}}
\put(7.5,1){\circle*{1.5}}
\put(14,1){\circle{2}}
\end{picture} }
\put(45,18){
\unitlength1.mm
\begin{picture}(15,2)
\linethickness{0.15mm}
\put(2,1){\vector(1,0){3.75}}
\put(7.5,1){\line(-1,0){1.75}}
\put(7.5,1){\vector(1,0){3.75}}
\put(13,1){\line(-1,0){1.75}}
\put(1,1){\circle{2}}
\put(7.5,1){\circle*{1.5}}
\put(14,1){\circle{2}}
\end{picture}  }
\put(100,18){
\unitlength1.mm
\begin{picture}(15,2)
\linethickness{0.15mm}
\put(13,1){\vector(-1,0){3.75}}
\put(7.5,1){\line(1,0){1.75}}
\put(7.5,1){\vector(-1,0){3.75}}
\put(2,1){\line(1,0){1.75}}
\put(14,1){\circle{2}}
\put(7.5,1){\circle*{1.5}}
\put(1,1){\circle{2}}
\end{picture}   }
\put(0,27.5){
\parbox[t]{15cm}{
\begin{eqnarray}
\label{S2}
=&&\ \ \ \; \hspace{4.67cm}+\nonumber\\[0.5cm]
=&&\ \ \ \ \psi_2(x+e_1)\ \bar\psi_2(x-e_1)\ \ \ \ \  +\ \ \ \ 
\ \psi_1(x-e_1)\ \bar\psi_1(x+e_1)\ \ \
\end{eqnarray}  }}
\end{picture} 

\vspace{7mm}

\unitlength1.mm
\begin{picture}(150,25)
\put(6.5,16.5){
\unitlength1.mm
\begin{picture}(8.5,8.5)
\linethickness{0.15mm}
\put(1,7.5){\line(1,0){5.5}}
\put(1,7.5){\line(0,-1){5.5}}
\put(7.5,7.5){\circle{2}}
\put(1,7.5){\circle*{1.5}}
\put(1,1){\circle{2}}
\end{picture}  }
\put(51.5,16.5){
\unitlength1.mm
\begin{picture}(8.5,8.5)
\linethickness{0.15mm}
\put(6.5,7.5){\vector(-1,0){3.75}}
\put(1,7.5){\line(1,0){1.75}}
\put(1,7.5){\vector(0,-1){3.75}}
\put(1,2){\line(0,1){1.75}}
\put(7.5,7.5){\circle{2}}
\put(1,7.5){\circle*{1.5}}
\put(1,1){\circle{2}}
\end{picture} }
\put(106.5,16.5){
\unitlength1.mm
\begin{picture}(8.5,8.5)
\linethickness{0.15mm}
\put(1,2){\vector(0,1){3.75}}
\put(1,7.5){\line(0,-1){1.75}}
\put(1,7.5){\vector(1,0){3.75}}
\put(6.5,7.5){\line(-1,0){1.75}}
\put(1,1){\circle{2}}
\put(1,7.5){\circle*{1.5}}
\put(7.5,7.5){\circle{2}}
\end{picture}  }
\put(0,33.5){
\parbox[t]{15cm}{
\begin{eqnarray}
\label{S3}
=&&\ \ \ \ {1\over\sqrt{2}}\hspace{4.1cm}-\ 
{1\over\sqrt{2}}\nonumber\\[11.5mm]
=&&\ \ \ \ \chi_1(x-e_2)\ \bar\psi_1(x+e_1)/\sqrt{2}\ -
\ \psi_2(x+e_1)\ \bar\chi_2(x-e_2)/\sqrt{2}
\end{eqnarray}  }}
\end{picture} 

\vspace{7mm}

\unitlength1.mm
\begin{picture}(150,25)
\put(0,16.5){
\unitlength1.mm
\begin{picture}(8.5,8.5)
\linethickness{0.15mm}
\put(7.5,1){\line(-1,0){5.5}}
\put(7.5,1){\line(0,1){5.5}}
\put(7.5,7.5){\circle{2}}
\put(7.5,1){\circle*{1.5}}
\put(1,1){\circle{2}}
\end{picture}  }
\put(45,16.5){
\unitlength1.mm
\begin{picture}(8.5,8.5)
\linethickness{0.15mm}
\put(7.5,6.5){\vector(0,-1){3.75}}
\put(7.5,1){\line(0,1){1.75}}
\put(7.5,1){\vector(-1,0){3.75}}
\put(2,1){\line(1,0){1.75}}
\put(7.5,7.5){\circle{2}}
\put(7.5,1){\circle*{1.5}}
\put(1,1){\circle{2}}
\end{picture} }
\put(100,16.5){
\unitlength1.mm
\begin{picture}(8.5,8.5)
\linethickness{0.15mm}
\put(2,1){\vector(1,0){3.75}}
\put(7.5,1){\line(-1,0){1.75}}
\put(7.5,1){\vector(0,1){3.75}}
\put(7.5,6.5){\line(0,-1){1.75}}
\put(1,1){\circle{2}}
\put(7.5,1){\circle*{1.5}}
\put(7.5,7.5){\circle{2}}
\end{picture}  }
\put(0,27){
\parbox[t]{15cm}{
\begin{eqnarray}
\label{S4}
=&&\ \ \ \ {1\over\sqrt{2}}\hspace{4.1cm}-\ {1\over\sqrt{2}}\nonumber\\[5mm]
=&&\ \ \ \ \psi_1(x-e_1)\ \bar\chi_1(x+e_2)/\sqrt{2}\ -
\ \chi_2(x+e_2)\ \bar\psi_2(x-e_1)/\sqrt{2}
\end{eqnarray}  }}
\end{picture} 

\vspace{7mm}

\unitlength1.mm
\begin{picture}(150,25)
\put(6.5,16.5){
\unitlength1.mm
\begin{picture}(8.5,8.5)
\linethickness{0.15mm}
\put(1,1){\line(0,1){5.5}}
\put(1,1){\line(1,0){5.5}}
\put(1,7.5){\circle{2}}
\put(1,1){\circle*{1.5}}
\put(7.5,1){\circle{2}}
\end{picture}  }
\put(51.5,16.5){
\unitlength1.mm
\begin{picture}(8.5,8.5)
\linethickness{0.15mm}
\put(1,6.5){\vector(0,-1){3.75}}
\put(1,1){\line(0,1){1.75}}
\put(1,1){\vector(1,0){3.75}}
\put(6.5,1){\line(-1,0){1.75}}
\put(1,7.5){\circle{2}}
\put(1,1){\circle*{1.5}}
\put(7.5,1){\circle{2}}
\end{picture}  }
\put(106.5,16.5){
\unitlength1.mm
\begin{picture}(8.5,8.5)
\linethickness{0.15mm}
\put(6.5,1){\vector(-1,0){3.75}}
\put(1,1){\line(1,0){1.75}}
\put(1,1){\vector(0,1){3.75}}
\put(1,6.5){\line(0,-1){1.75}}
\put(7.5,1){\circle{2}}
\put(1,1){\circle*{1.5}}
\put(1,7.5){\circle{2}}
\end{picture}  }
\put(0,27){
\parbox[t]{15cm}{
\begin{eqnarray}
\label{S5}
=&&\ \ \ \ {1\over\sqrt{2}}\hspace{4.1cm}+\ {1\over\sqrt{2}}\nonumber\\[5mm]
=&&\ \ \ \ \psi_2(x+e_1)\ \bar\chi_1(x+e_2)/\sqrt{2} \ +
\ \chi_2(x+e_2)\ \bar\psi_1(x+e_1)/\sqrt{2}
\end{eqnarray}  }}
\end{picture} 

\vspace{7mm}

\unitlength1.mm
\begin{picture}(150,25)
\put(0,16.5){
\unitlength1.mm
\begin{picture}(8.5,8.5)
\linethickness{0.15mm}
\put(7.5,7.5){\line(0,-1){5.5}}
\put(7.5,7.5){\line(-1,0){5.5}}
\put(1,7.5){\circle{2}}
\put(7.5,7.5){\circle*{1.5}}
\put(7.5,1){\circle{2}}
\end{picture}  }
\put(45,16.5){
\unitlength1.mm
\begin{picture}(8.5,8.5)
\linethickness{0.15mm}
\put(2,7.5){\vector(1,0){3.75}}
\put(7.5,7.5){\line(-1,0){1.75}}
\put(7.5,7.5){\vector(0,-1){3.75}}
\put(7.5,2){\line(0,1){1.75}}
\put(1,7.5){\circle{2}}
\put(7.5,7.5){\circle*{1.5}}
\put(7.5,1){\circle{2}}
\end{picture}  }
\put(100,16.5){
\unitlength1.mm
\begin{picture}(2,8)
\linethickness{0.15mm}
\put(7.5,2){\vector(0,1){3.75}}
\put(7.5,7.5){\line(0,-1){1.75}}
\put(7.5,7.5){\vector(-1,0){3.75}}
\put(2,7.5){\line(1,0){1.75}}
\put(1,7.5){\circle{2}}
\put(7.5,7.5){\circle*{1.5}}
\put(7.5,1){\circle{2}}
\end{picture}  }
\put(0,33.5){
\parbox[t]{15cm}{
\begin{eqnarray}
\label{S6}
=&&\ \ \ \ {1\over\sqrt{2}}\hspace{4.1cm}+\ 
{1\over\sqrt{2}}\nonumber\\[11.5mm]
=&&\ \ \ \ \chi_1(x-e_2)\ \bar\psi_2(x-e_1)/\sqrt{2} \ + 
\ \psi_1(x-e_1)\ \bar\chi_2(x-e_2)/\sqrt{2}
\end{eqnarray}  }}
\end{picture}

One immediately recognizes the significance of the numerical 
factors for each vertex. The rule simply is that each corner
element (\ref{S3})-(\ref{S6}) is associated with a factor $1/\sqrt{2}$.
Using the above pictorial language for the remaining fields and 
their combinations on the odd sublattice we can gain now
already some intuitive understanding for the structures 
appearing on the r.h.s.\ of eqs.\ (\ref{V2})-(\ref{V20})
by interpreting them in terms of the graphical symbols
introduced in eqs.\ (\ref{F1})-(\ref{F8}), (\ref{S1})-(\ref{S6}). 
We give some characteristic examples; similar relations can
also be obtained for the omitted vertices.\\

Vertex 3
\nopagebreak

\vspace{5mm}
\nopagebreak

\unitlength1.mm
\begin{picture}(150,15)
\put(0,0){
\unitlength1.mm
\begin{picture}(15,15)
\linethickness{0.15mm}
\put(7.5,12){\line(0,1){1}}
\put(7.5,10){\line(0,1){1}}
\put(7.5, 8){\line(0,1){1}}
\put(7.5, 6){\line(0,1){1}}
\put(7.5, 4){\line(0,1){1}}
\put(7.5, 2){\line(0,1){1}}
\linethickness{0.6mm}
\put(2,7.5){\line(1,0){11}}
\put(7.5,7.5){\circle*{1.5}}
\end{picture} }
\put(38,6.5){
\unitlength1.mm
\begin{picture}(15,2)
\linethickness{0.15mm}
\put(2,1){\line(1,0){11}}
\put(1,1){\circle{2}}
\put(7.5,1){\circle*{1.5}}
\put(14,1){\circle{2}}
\end{picture} }
\put(63,6.5){
\unitlength1.mm
\begin{picture}(15,2)
\linethickness{0.15mm}
\put(2,1){\line(1,0){11}}
\put(1,1){\circle{2}}
\put(7.5,1){\circle*{1.5}}
\put(14,1){\circle{2}}
\end{picture} }
\put(95.25,6.5){
\unitlength1.mm
\begin{picture}(15,2)
\linethickness{0.15mm}
\put(2,1){\vector(1,0){3.75}}
\put(7.5,1){\line(-1,0){1.75}}
\put(7.5,1){\vector(1,0){3.75}}
\put(13,1){\line(-1,0){1.75}}
\put(1,1){\circle{2}}
\put(7.5,1){\circle*{1.5}}
\put(14,1){\circle{2}}
\end{picture}  }
\put(120.25,6.5){
\unitlength1.mm
\begin{picture}(15,2)
\linethickness{0.15mm}
\put(13,1){\vector(-1,0){3.75}}
\put(7.5,1){\line(1,0){1.75}}
\put(7.5,1){\vector(-1,0){3.75}}
\put(2,1){\line(1,0){1.75}}
\put(14,1){\circle{2}}
\put(7.5,1){\circle*{1.5}}
\put(1,1){\circle{2}}
\end{picture}   }
\put(0,16.5){
\parbox[t]{15cm}{
\begin{eqnarray}
\label{R1}
&&=\ \ {1\over 2}
\hspace{2.6cm}\times\hspace{2.45cm}=\ \ \
\hspace{2cm}\times\hspace{1.85cm}
\end{eqnarray}  }}
\end{picture} 

\vspace{3mm}

Vertex 6
\nopagebreak

\vspace{5mm}
\nopagebreak

\unitlength1.mm
\begin{picture}(150,15)
\put(0,0){
\unitlength1.mm
\begin{picture}(15,15)
\linethickness{0.15mm}
\put( 6,7.5){\line(1,0){1}}
\put( 4,7.5){\line(1,0){1}}
\put( 2,7.5){\line(1,0){1}}
\put(7.5, 6){\line(0,1){1}}
\put(7.5, 4){\line(0,1){1}}
\put(7.5, 2){\line(0,1){1}}
\linethickness{0.6mm}
\put(7.5,7.5){\line(1,0){5.5}}
\put(7.5,7.5){\line(0,1){5.5}}
\put(7.5,7.5){\circle*{1.5}}
\end{picture} }
\put(44.5,6.5){
\unitlength1.mm
\begin{picture}(8.5,8.5)
\linethickness{0.15mm}
\put(1,1){\line(0,1){5.5}}
\put(1,1){\line(1,0){5.5}}
\put(1,7.5){\circle{2}}
\put(1,1){\circle*{1.5}}
\put(7.5,1){\circle{2}}
\end{picture}  }
\put(69.5,6.5){
\unitlength1.mm
\begin{picture}(8.5,8.5)
\linethickness{0.15mm}
\put(1,1){\line(0,1){5.5}}
\put(1,1){\line(1,0){5.5}}
\put(1,7.5){\circle{2}}
\put(1,1){\circle*{1.5}}
\put(7.5,1){\circle{2}}
\end{picture}  }
\put(101.75,6.5){
\unitlength1.mm
\begin{picture}(8.5,8.5)
\linethickness{0.15mm}
\put(1,6.5){\vector(0,-1){3.75}}
\put(1,1){\line(0,1){1.75}}
\put(1,1){\vector(1,0){3.75}}
\put(6.5,1){\line(-1,0){1.75}}
\put(1,7.5){\circle{2}}
\put(1,1){\circle*{1.5}}
\put(7.5,1){\circle{2}}
\end{picture}  }
\put(126.75,6.5){
\unitlength1.mm
\begin{picture}(8.5,8.5)
\linethickness{0.15mm}
\put(6.5,1){\vector(-1,0){3.75}}
\put(1,1){\line(1,0){1.75}}
\put(1,1){\vector(0,1){3.75}}
\put(1,6.5){\line(0,-1){1.75}}
\put(7.5,1){\circle{2}}
\put(1,1){\circle*{1.5}}
\put(1,7.5){\circle{2}}
\end{picture}  }
\put(0,16.5){
\parbox[t]{15cm}{
\begin{eqnarray}
\label{R2}
&&=\ \ {1\over 2}
\hspace{2.6cm}\times\hspace{2.45cm}=\ \, {1\over 2}
\hspace{2cm}\times\hspace{1.85cm}
\end{eqnarray}  }}
\end{picture}

\vspace{3mm} 

Since the vertices 8-20 correspond to sums of field products 
the following condensed notation is appropriate.
Each of the pictures below however can also be expanded in terms
of oriented line segments.\\

Vertex 10
\nopagebreak

\vspace{5mm}
\nopagebreak

\unitlength1.mm
\begin{picture}(150,15)
\put(0,0){
\unitlength1.mm
\begin{picture}(15,15)
\linethickness{0.15mm}
\put(2,7.5){\line(1,0){11}}
\put(7.5, 6){\line(0,1){1}}
\put(7.5, 4){\line(0,1){1}}
\put(7.5, 2){\line(0,1){1}}
\linethickness{0.6mm}
\put(7.5,7.5){\line(0,1){5.5}}
\put(7.5,7.5){\circle*{1.5}}
\end{picture}  }
\put(38,6.5){
\unitlength1.mm
\begin{picture}(8.5,8.5)
\linethickness{0.15mm}
\put(7.5,1){\line(-1,0){5.5}}
\put(7.5,1){\line(0,1){5.5}}
\put(7.5,7.5){\circle{2}}
\put(7.5,1){\circle*{1.5}}
\put(1,1){\circle{2}}
\end{picture} }
\put(69.5,6.5){
\unitlength1.mm
\begin{picture}(8.5,8.5)
\linethickness{0.15mm}
\put(1,1){\line(0,1){5.5}}
\put(1,1){\line(1,0){5.5}}
\put(1,7.5){\circle{2}}
\put(1,1){\circle*{1.5}}
\put(7.5,1){\circle{2}}
\end{picture} }
\put(0,16){
\parbox[t]{15cm}{
\begin{eqnarray}
\label{R3}
&&=\ \ \ 
\hspace{2.65cm}\times\hspace{7.6cm}
\end{eqnarray}  }}
\end{picture}

\vspace{3mm} 

Vertex 14
\nopagebreak

\vspace{5mm}
\nopagebreak

\unitlength1.mm
\begin{picture}(150,15)
\put(0,0){
\unitlength1.mm
\begin{picture}(15,15)
\linethickness{0.15mm}
\put(7.5, 6){\line(0,1){1}}
\put(7.5, 4){\line(0,1){1}}
\put(7.5, 2){\line(0,1){1}}
\put(2,7.5){\line(1,0){5.5}}
\put(7.5,7.5){\line(0,1){5.5}}
\linethickness{0.6mm}
\put(7.5,7.5){\line(1,0){5.5}}
\put(7.5,7.5){\circle*{1.5}}
\end{picture}  }
\put(38,6.5){
\unitlength1.mm
\begin{picture}(15,2)
\linethickness{0.15mm}
\put(2,1){\line(1,0){11}}
\put(1,1){\circle{2}}
\put(7.5,1){\circle*{1.5}}
\put(14,1){\circle{2}}
\end{picture}  }
\put(69.5,6.5){
\unitlength1.mm
\begin{picture}(8.5,8.5)
\linethickness{0.15mm}
\put(1,1){\line(0,1){5.5}}
\put(1,1){\line(1,0){5.5}}
\put(1,7.5){\circle{2}}
\put(1,1){\circle*{1.5}}
\put(7.5,1){\circle{2}}
\end{picture} }
\put(0,16){
\parbox[t]{15cm}{
\begin{eqnarray}
\label{R4}
&&=\ \ \ 
\hspace{2.65cm}\times\hspace{7.6cm}
\end{eqnarray}  }}
\end{picture}

\vspace{3mm} 

Vertex 20
\nopagebreak

\vspace{5mm}
\nopagebreak

\unitlength1.mm
\begin{picture}(150,35)
\put(0,20){
\unitlength1.mm
\begin{picture}(15,15)
\linethickness{0.15mm}
\put(2,7.5){\line(1,0){11}}
\put(7.5,2){\line(0,1){11}}
\put(7.5,7.5){\circle*{1.5}}
\end{picture} }
\put(38,26.5){
\unitlength1.mm
\begin{picture}(8.5,8.5)
\linethickness{0.15mm}
\put(7.5,1){\line(-1,0){5.5}}
\put(7.5,1){\line(0,1){5.5}}
\put(7.5,7.5){\circle{2}}
\put(7.5,1){\circle*{1.5}}
\put(1,1){\circle{2}}
\end{picture} }
\put(69.5,20){
\unitlength1.mm
\begin{picture}(8.5,8.5)
\linethickness{0.15mm}
\put(1,7.5){\line(1,0){5.5}}
\put(1,7.5){\line(0,-1){5.5}}
\put(7.5,7.5){\circle{2}}
\put(1,7.5){\circle*{1.5}}
\put(1,1){\circle{2}}
\end{picture}  }
\put(95.25,20){
\unitlength1.mm
\begin{picture}(8.5,8.5)
\linethickness{0.15mm}
\put(7.5,7.5){\line(0,-1){5.5}}
\put(7.5,7.5){\line(-1,0){5.5}}
\put(1,7.5){\circle{2}}
\put(7.5,7.5){\circle*{1.5}}
\put(7.5,1){\circle{2}}
\end{picture}  }
\put(126.75,26.5){
\unitlength1.mm
\begin{picture}(8.5,8.5)
\linethickness{0.15mm}
\put(1,1){\line(0,1){5.5}}
\put(1,1){\line(1,0){5.5}}
\put(1,7.5){\circle{2}}
\put(1,1){\circle*{1.5}}
\put(7.5,1){\circle{2}}
\end{picture} }
\put(44.5,0){
\unitlength1.mm
\begin{picture}(2,15)
\linethickness{0.15mm}
\put(1,2){\line(0,1){11}}
\put(1,14){\circle{2}}
\put(1,7.5){\circle*{1.5}}
\put(1,1){\circle{2}}
\end{picture}  }
\put(63,6.5){
\unitlength1.mm
\begin{picture}(15,2)
\linethickness{0.15mm}
\put(2,1){\line(1,0){11}}
\put(1,1){\circle{2}}
\put(7.5,1){\circle*{1.5}}
\put(14,1){\circle{2}}
\end{picture} }
\put(0,36){
\parbox[t]{15cm}{
\begin{eqnarray}
\label{R5}
&&=\ \ \
\hspace{2.6cm}\times\hspace{2.5cm}+\ \ \
\hspace{2cm}\times\hspace{1.85cm}\nonumber\\[1.4cm]
&&\ \ \ \ + 
\hspace{2.6cm}\times\hspace{2.6cm}\ \ \ \
\hspace{1.9cm}\ \hspace{1.85cm}
\end{eqnarray}  }}
\end{picture}

\vspace{5mm}

It should be mentioned that most of the above relations 
(\ref{R1})-(\ref{R5}) do not generalize
to the case when the Wilson parameter $r$ is chosen not to be 1.
However, interestingly enough eq.\ (\ref{R5}) still applies for
$r\neq 1$ if for the graphical symbols (\ref{S1})-(\ref{S6})
the expressions then obtained for the vertices 21-26 (eqs.\ 
(\ref{V21})-(\ref{V26})) are inserted
(with $M = 1$).\\

\parindent1.5em

What remains to be done now is to discuss the result for the 
Grassmann integration on the odd sublattice $\Lambda_o$. 
For the moment we 
exclude vertex 20 (and its equivalent on the odd sublattice) from 
consideration as its complexity requires special 
attention. It will be discussed at a later stage, in subsection
2.2. Now, it is useful to introduce a graphical notation
for the product of two field variables at a certain point $x$ of the 
odd sublattice $\Lambda_o$. In view of the vertex arrow rule
we only need products of two Grassmann fields 
where one of them carries a bar (i.e.\ oriented line segments
in the pictorial language). We order the products always in such 
a way that for a given lattice point the out-arrow field variable
stands first (For points on the odd sublattice
this entails that the bar variable is the first factor in a product while 
for points on the even lattice it is always the second factor;
see e.g.\ the ordering applied in eqs.\ (\ref{V2})-(\ref{V26}).).
We give just an example; 
all other combinations are defined in an analogous way (see eqs.\
(\ref{F1})-(\ref{F8}) for the notation). 

\vspace{7mm}

\parindent0.em

\unitlength1.mm
\begin{picture}(150,7)
\put(0,5){
\unitlength1.mm
\begin{picture}(8.25,2)
\linethickness{0.15mm}
\put(6.25,1){\vector(-1,0){3.75}}
\put(0.75,1){\line(1,0){1.75}}
\put(7.25,1){\circle{2}}
\put(0.75,1){\circle*{1.5}}
\end{picture}  }
\put(6.25,5){
\unitlength1.mm
\begin{picture}(8.25,2)
\linethickness{0.15mm}
\put(7.5,1){\vector(-1,0){3.75}}
\put(2,1){\line(1,0){1.75}}
\put(1,1){\circle{2}}
\put(7.5,1){\circle*{1.5}}
\end{picture}   }
\put(0,14.5){
\parbox[t]{15cm}{
\begin{eqnarray}
\label{F9}
&&=\ \bar\psi_1(x)\ \psi_1(x)\hspace{9cm}
\end{eqnarray}  }}
\end{picture} 

It is also useful to establish a graphical symbol for the mass
term combination of the fields.

\vspace{7mm}

\unitlength1.mm
\begin{picture}(150,7)
\put(5,5){
\unitlength1.mm
\begin{picture}(4,4)
\linethickness{0.15mm}
\put(2,2){\circle{4}}
\end{picture}   }
\put(0,15){
\parbox[t]{15cm}{
\begin{eqnarray}
\label{F10}
&&=\ M\left[ \bar\psi_1(x)\ \psi_1(x)\ +\ \bar\psi_2(x)\ \psi_2(x)\right]
\hspace{6cm}
\end{eqnarray}  }}
\end{picture} 

Using these symbols we can now conveniently discuss the 
Grassmann integration on the odd sublattice $\Lambda_o$.
The graphical symbols below denote in an intuitive way 
the product of (four) Grassmann variables (and sums of them
in eqs.\ (\ref{O7})-(\ref{O9})). 
The ordering is performed within pairs of fields which 
stand for an oriented line segment. In each pair the bar variable
stands first. As these products of two Grassmann variables are
Grassmann even, no further ordering prescription is required.
The black dot in the hollow circle of the point on 
the odd sublattice $\Lambda_o$ (which is the central point in
the diagrams below)
indicates that the Grassmann integration has now been performed.
In view of the vertex arrow rule one finds the following non-zero 
results (Dots which are close together denote one and the same 
lattice point). We give some characteristic examples only. 

\vspace{7mm}

\unitlength1.mm
\begin{picture}(150,9.5)
\put(0,5){
\unitlength1.mm
\begin{picture}(8.25,2)
\linethickness{0.15mm}
\put(6.25,1){\vector(-1,0){3.75}}
\put(0.75,1){\line(1,0){1.75}}
\put(7.25,1){\circle{2}}
\put(0.75,1){\circle*{1.5}}
\end{picture}  }
\put(6.25,5){
\unitlength1.mm
\begin{picture}(8.25,2)
\linethickness{0.15mm}
\put(7.5,1){\vector(-1,0){3.75}}
\put(2,1){\line(1,0){1.75}}
\put(1,1){\circle*{1.5}}
\put(7.5,1){\circle*{1.5}}
\end{picture}   }
\put(0,6.5){
\unitlength1.mm
\begin{picture}(8.25,2)
\linethickness{0.15mm}
\put(0.75,1){\vector(1,0){3.75}}
\put(6.25,1){\line(-1,0){1.75}}
\put(7.25,1){\circle{2}}
\put(0.75,1){\circle*{1.5}}
\end{picture}   }
\put(6.25,6.5){
\unitlength1.mm
\begin{picture}(8.25,2)
\linethickness{0.15mm}
\put(2,1){\vector(1,0){3.75}}
\put(7.5,1){\line(-1,0){1.75}}
\put(1,1){\circle*{1.5}}
\put(7.5,1){\circle*{1.5}}
\end{picture}  }
\put(0,17){
\parbox[t]{15cm}{
\begin{eqnarray}
\label{O1}
&&=\ \int \prod_{\alpha=1}^2
d\psi_\alpha(x)\ d\bar{\psi}_\alpha(x)\ 
\bar{\psi}_1(x)\ \psi_1(x)\ \bar{\psi}_2(x)\ \psi_2(x)\ =
\ \ 1\hspace{1cm}
\end{eqnarray}  }}
\end{picture} 

\vspace{7mm}

\unitlength1.mm
\begin{picture}(150,9.5)
\put(6.25,5){
\unitlength1.mm
\begin{picture}(2,8.25)
\linethickness{0.15mm}
\put(1,7.5){\vector(0,-1){3.75}}
\put(1,2){\line(0,1){1.75}}
\put(1,1){\circle{2}}
\put(1,7.5){\circle*{1.5}}
\end{picture}    }
\put(0,5){
\unitlength1.mm
\begin{picture}(8.25,2)
\linethickness{0.15mm}
\put(6.25,1){\vector(-1,0){3.75}}
\put(0.75,1){\line(1,0){1.75}}
\put(7.25,1){\circle*{1.5}}
\put(0.75,1){\circle*{1.5}}
\end{picture}  }
\put(0,6.5){
\unitlength1.mm
\begin{picture}(8.25,2)
\linethickness{0.15mm}
\put(0.75,1){\vector(1,0){3.75}}
\put(4.75,1){\line(-1,0){0.25}}
\put(5.75,1){\circle{2}}
\put(0.75,1){\circle*{1.5}}
\end{picture}   }
\put(4.75,5){
\unitlength1.mm
\begin{picture}(2,8.25)
\linethickness{0.15mm}
\put(1,3.5){\vector(0,1){2.25}}
\put(1,7.5){\line(0,-1){1.75}}
\put(1,2.5){\circle*{1.5}}
\put(1,7.5){\circle*{1.5}}
\end{picture}  }
\put(0,16){
\parbox[t]{15cm}{
\begin{eqnarray}
\label{O2}
&&=\ -\ {1\over 2}\hspace{9.5cm}
\end{eqnarray}  }}
\end{picture} 

\vspace{7mm}

\unitlength1.mm
\begin{picture}(150,9.5)
\put(6.25,5){
\unitlength1.mm
\begin{picture}(2,8.25)
\linethickness{0.15mm}
\put(1,7.5){\vector(0,-1){3.75}}
\put(1,2){\line(0,1){1.75}}
\put(1,1){\circle{2}}
\put(1,7.5){\circle*{1.5}}
\end{picture}    }
\put(6.25,5){
\unitlength1.mm
\begin{picture}(8.25,2)
\linethickness{0.15mm}
\put(2,1){\vector(1,0){3.75}}
\put(7.5,1){\line(-1,0){1.75}}
\put(1,1){\circle*{1.5}}
\put(7.5,1){\circle*{1.5}}
\end{picture}  }
\put(6.25,6.5){
\unitlength1.mm
\begin{picture}(8.25,2)
\linethickness{0.15mm}
\put(7.5,1){\vector(-1,0){3.75}}
\put(3.5,1){\line(1,0){0.25}}
\put(2.5,1){\circle{2}}
\put(7.5,1){\circle*{1.5}}
\end{picture}   }
\put(7.75,5){
\unitlength1.mm
\begin{picture}(2,8.25)
\linethickness{0.15mm}
\put(1,3.5){\vector(0,1){2.25}}
\put(1,7.5){\line(0,-1){1.75}}
\put(1,2.5){\circle*{1.5}}
\put(1,7.5){\circle*{1.5}}
\end{picture}  }
\put(0,16){
\parbox[t]{15cm}{
\begin{eqnarray}
\label{O3}
&&=\ \ \ \ \; {1\over 2}\hspace{9.5cm}
\end{eqnarray}  }}
\end{picture} 

\vspace{7mm}

\unitlength1.mm
\begin{picture}(150,9.5)
\put(6.25,5){
\unitlength1.mm
\begin{picture}(2,8.25)
\linethickness{0.15mm}
\put(1,7.5){\vector(0,-1){3.75}}
\put(1,2){\line(0,1){1.75}}
\put(1,1){\circle{2}}
\put(1,7.5){\circle*{1.5}}
\end{picture}    }
\put(0,5){
\unitlength1.mm
\begin{picture}(8.25,2)
\linethickness{0.15mm}
\put(6.25,1){\vector(-1,0){3.75}}
\put(0.75,1){\line(1,0){1.75}}
\put(7.25,1){\circle*{1.5}}
\put(0.75,1){\circle*{1.5}}
\end{picture}  }
\put(7.75,5){
\unitlength1.mm
\begin{picture}(8.25,2)
\linethickness{0.15mm}
\put(7.5,1){\vector(-1,0){3.75}}
\put(2,1){\line(1,0){1.75}}
\put(1,1){\circle{2}}
\put(7.5,1){\circle*{1.5}}
\end{picture}   }
\put(7.75,5){
\unitlength1.mm
\begin{picture}(2,8.25)
\linethickness{0.15mm}
\put(1,2){\vector(0,1){3.75}}
\put(1,7.5){\line(0,-1){1.75}}
\put(1,1){\circle*{1.5}}
\put(1,7.5){\circle*{1.5}}
\end{picture}  }
\put(45,5){
\unitlength1.mm
\begin{picture}(8.25,2)
\linethickness{0.15mm}
\put(0.75,1){\vector(1,0){3.75}}
\put(6.25,1){\line(-1,0){1.75}}
\put(7.25,1){\circle{2}}
\put(0.75,1){\circle*{1.5}}
\end{picture}   }
\put(51.25,5){
\unitlength1.mm
\begin{picture}(2,8.25)
\linethickness{0.15mm}
\put(1,2){\vector(0,1){3.75}}
\put(1,7.5){\line(0,-1){1.75}}
\put(1,1){\circle*{1.5}}
\put(1,7.5){\circle*{1.5}}
\end{picture}  }
\put(52.75,5){
\unitlength1.mm
\begin{picture}(2,8.25)
\linethickness{0.15mm}
\put(1,7.5){\vector(0,-1){3.75}}
\put(1,2){\line(0,1){1.75}}
\put(1,1){\circle{2}}
\put(1,7.5){\circle*{1.5}}
\end{picture}    }
\put(52.75,5){
\unitlength1.mm
\begin{picture}(8.25,2)
\linethickness{0.15mm}
\put(2,1){\vector(1,0){3.75}}
\put(7.5,1){\line(-1,0){1.75}}
\put(1,1){\circle*{1.5}}
\put(7.5,1){\circle*{1.5}}
\end{picture}  }
\put(0,15){
\parbox[t]{15cm}{
\begin{eqnarray}
\label{O4}
&&=\ -\hspace{3.5cm}=\ {1\over2}\hspace{5.5cm}
\end{eqnarray}  }}
\end{picture} 

\vspace{7mm}

\unitlength1.mm
\begin{picture}(150,9.5)
\put(0.75,5){
\unitlength1.mm
\begin{picture}(8.25,2)
\linethickness{0.15mm}
\put(0.75,1){\vector(1,0){3.75}}
\put(6.25,1){\line(-1,0){1.75}}
\put(7.25,1){\circle{2}}
\put(0.75,1){\circle*{1.5}}
\end{picture}   }
\put(7,5){
\unitlength1.mm
\begin{picture}(8.25,2)
\linethickness{0.15mm}
\put(2,1){\vector(1,0){3.75}}
\put(7.5,1){\line(-1,0){1.75}}
\put(1,1){\circle*{1.5}}
\put(7.5,1){\circle*{1.5}}
\end{picture}  }
\put(7,6.5){
\unitlength1.mm
\begin{picture}(8.25,2)
\linethickness{0.15mm}
\put(7.5,1){\vector(-1,0){3.75}}
\put(2,1){\line(1,0){1.75}}
\put(1,1){\circle{2}}
\put(7.5,1){\circle*{1.5}}
\end{picture}   }
\put(7,6.5){
\unitlength1.mm
\begin{picture}(2,8.25)
\linethickness{0.15mm}
\put(1,2){\vector(0,1){3.75}}
\put(1,7.5){\line(0,-1){1.75}}
\put(1,1){\circle*{1.5}}
\put(1,7.5){\circle*{1.5}}
\end{picture}  }
\put(52,5){
\unitlength1.mm
\begin{picture}(8.25,2)
\linethickness{0.15mm}
\put(7.5,1){\vector(-1,0){3.75}}
\put(2,1){\line(1,0){1.75}}
\put(1,1){\circle{2}}
\put(7.5,1){\circle*{1.5}}
\end{picture}   }
\put(45.75,5){
\unitlength1.mm
\begin{picture}(8.25,2)
\linethickness{0.15mm}
\put(6.25,1){\vector(-1,0){3.75}}
\put(0.75,1){\line(1,0){1.75}}
\put(7.25,1){\circle*{1.5}}
\put(0.75,1){\circle*{1.5}}
\end{picture}  }
\put(52,6.5){
\unitlength1.mm
\begin{picture}(2,8.25)
\linethickness{0.15mm}
\put(1,7.5){\vector(0,-1){3.75}}
\put(1,2){\line(0,1){1.75}}
\put(1,1){\circle{2}}
\put(1,7.5){\circle*{1.5}}
\end{picture}    }
\put(52,6.5){
\unitlength1.mm
\begin{picture}(8.25,2)
\linethickness{0.15mm}
\put(2,1){\vector(1,0){3.75}}
\put(7.5,1){\line(-1,0){1.75}}
\put(1,1){\circle*{1.5}}
\put(7.5,1){\circle*{1.5}}
\end{picture}  }
\put(0,16){
\parbox[t]{15cm}{
\begin{eqnarray}
\label{O5}
&&=\ \ \ \ \ \hspace{3.5cm}=\ {1\over\sqrt{2}}\hspace{5.1cm}
\end{eqnarray}  }}
\end{picture} 

\vspace{7mm}

\unitlength1.mm
\begin{picture}(150,9.5)
\put(0.75,5){
\unitlength1.mm
\begin{picture}(8.25,2)
\linethickness{0.15mm}
\put(0.75,1){\vector(1,0){3.75}}
\put(6.25,1){\line(-1,0){1.75}}
\put(7.25,1){\circle{2}}
\put(0.75,1){\circle*{1.5}}
\end{picture}   }
\put(7,5){
\unitlength1.mm
\begin{picture}(8.25,2)
\linethickness{0.15mm}
\put(2,1){\vector(1,0){3.75}}
\put(7.5,1){\line(-1,0){1.75}}
\put(1,1){\circle*{1.5}}
\put(7.5,1){\circle*{1.5}}
\end{picture}  }
\put(7,6.5){
\unitlength1.mm
\begin{picture}(2,8.25)
\linethickness{0.15mm}
\put(1,7.5){\vector(0,-1){3.75}}
\put(1,2){\line(0,1){1.75}}
\put(1,1){\circle{2}}
\put(1,7.5){\circle*{1.5}}
\end{picture}    }
\put(0.75,6.5){
\unitlength1.mm
\begin{picture}(8.25,2)
\linethickness{0.15mm}
\put(6.25,1){\vector(-1,0){3.75}}
\put(0.75,1){\line(1,0){1.75}}
\put(7.25,1){\circle*{1.5}}
\put(0.75,1){\circle*{1.5}}
\end{picture}  }
\put(52,5){
\unitlength1.mm
\begin{picture}(8.25,2)
\linethickness{0.15mm}
\put(7.5,1){\vector(-1,0){3.75}}
\put(2,1){\line(1,0){1.75}}
\put(1,1){\circle{2}}
\put(7.5,1){\circle*{1.5}}
\end{picture}   }
\put(45.75,5){
\unitlength1.mm
\begin{picture}(8.25,2)
\linethickness{0.15mm}
\put(6.25,1){\vector(-1,0){3.75}}
\put(0.75,1){\line(1,0){1.75}}
\put(7.25,1){\circle*{1.5}}
\put(0.75,1){\circle*{1.5}}
\end{picture}  }
\put(45.75,6.5){
\unitlength1.mm
\begin{picture}(8.25,2)
\linethickness{0.15mm}
\put(0.75,1){\vector(1,0){3.75}}
\put(6.25,1){\line(-1,0){1.75}}
\put(7.25,1){\circle{2}}
\put(0.75,1){\circle*{1.5}}
\end{picture}   }
\put(52,6.5){
\unitlength1.mm
\begin{picture}(2,8.25)
\linethickness{0.15mm}
\put(1,2){\vector(0,1){3.75}}
\put(1,7.5){\line(0,-1){1.75}}
\put(1,1){\circle*{1.5}}
\put(1,7.5){\circle*{1.5}}
\end{picture}  }
\put(0,16){
\parbox[t]{15cm}{
\begin{eqnarray}
\label{O6}
&&=\ -\hspace{3.5cm}=\ {1\over\sqrt{2}}\hspace{5.1cm}
\end{eqnarray}  }}
\end{picture} 

\vspace{7mm}

\unitlength1.mm
\begin{picture}(150,9.5)
\put(0.75,5){
\unitlength1.mm
\begin{picture}(8.25,2)
\linethickness{0.15mm}
\put(0.75,1){\vector(1,0){3.75}}
\put(6.25,1){\line(-1,0){1.75}}
\put(7.25,1){\circle{2}}
\put(7.25,1){\circle{4}}
\put(0.75,1){\circle*{1.5}}
\end{picture}   }
\put(7,5){
\unitlength1.mm
\begin{picture}(8.25,2)
\linethickness{0.15mm}
\put(2,1){\vector(1,0){3.75}}
\put(7.5,1){\line(-1,0){1.75}}
\put(1,1){\circle*{1.5}}
\put(7.5,1){\circle*{1.5}}
\end{picture}  }
\put(52,5){
\unitlength1.mm
\begin{picture}(8.25,2)
\linethickness{0.15mm}
\put(7.5,1){\vector(-1,0){3.75}}
\put(2,1){\line(1,0){1.75}}
\put(1,1){\circle{2}}
\put(1,1){\circle{4}}
\put(7.5,1){\circle*{1.5}}
\end{picture}   }
\put(45.75,5){
\unitlength1.mm
\begin{picture}(8.25,2)
\linethickness{0.15mm}
\put(6.25,1){\vector(-1,0){3.75}}
\put(0.75,1){\line(1,0){1.75}}
\put(7.25,1){\circle*{1.5}}
\put(0.75,1){\circle*{1.5}}
\end{picture}  }
\put(0,14.5){
\parbox[t]{15cm}{
\begin{eqnarray}
\label{O7}
&&=\ \ \ \ \ \hspace{3.5cm}=\ M\hspace{5.3cm}
\end{eqnarray}  }}
\end{picture} 

\vspace{7mm}

\unitlength1.mm
\begin{picture}(150,9.5)
\put(7,5){
\unitlength1.mm
\begin{picture}(2,8.25)
\linethickness{0.15mm}
\put(1,7.5){\vector(0,-1){3.75}}
\put(1,2){\line(0,1){1.75}}
\put(1,1){\circle{2}}
\put(1,1){\circle{4}}
\put(1,7.5){\circle*{1.5}}
\end{picture}    }
\put(0.75,5){
\unitlength1.mm
\begin{picture}(8.25,2)
\linethickness{0.15mm}
\put(6.25,1){\vector(-1,0){3.75}}
\put(0.75,1){\line(1,0){1.75}}
\put(7.25,1){\circle*{1.5}}
\put(0.75,1){\circle*{1.5}}
\end{picture}  }
\put(45.75,5){
\unitlength1.mm
\begin{picture}(8.25,2)
\linethickness{0.15mm}
\put(0.75,1){\vector(1,0){3.75}}
\put(6.25,1){\line(-1,0){1.75}}
\put(7.25,1){\circle{2}}
\put(7.25,1){\circle{4}}
\put(0.75,1){\circle*{1.5}}
\end{picture}   }
\put(52,5){
\unitlength1.mm
\begin{picture}(2,8.25)
\linethickness{0.15mm}
\put(1,2){\vector(0,1){3.75}}
\put(1,7.5){\line(0,-1){1.75}}
\put(1,1){\circle*{1.5}}
\put(1,7.5){\circle*{1.5}}
\end{picture}  }
\put(0,15.5){
\parbox[t]{15cm}{
\begin{eqnarray}
\label{O8}
&&=\ -\hspace{3.5cm}=\ {M\over\sqrt{2}}\hspace{5.1cm}
\end{eqnarray}  }}
\end{picture} 

\vspace{7mm}

\unitlength1.mm
\begin{picture}(150,9.5)
\put(7,5){
\unitlength1.mm
\begin{picture}(8.25,2)
\linethickness{0.15mm}
\put(7.5,1){\vector(-1,0){3.75}}
\put(2,1){\line(1,0){1.75}}
\put(1,1){\circle{2}}
\put(1,1){\circle{4}}
\put(7.5,1){\circle*{1.5}}
\end{picture}   }
\put(7,5){
\unitlength1.mm
\begin{picture}(2,8.25)
\linethickness{0.15mm}
\put(1,2){\vector(0,1){3.75}}
\put(1,7.5){\line(0,-1){1.75}}
\put(1,1){\circle*{1.5}}
\put(1,7.5){\circle*{1.5}}
\end{picture}  }
\put(52,5){
\unitlength1.mm
\begin{picture}(2,8.25)
\linethickness{0.15mm}
\put(1,7.5){\vector(0,-1){3.75}}
\put(1,2){\line(0,1){1.75}}
\put(1,1){\circle{2}}
\put(1,1){\circle{4}}
\put(1,7.5){\circle*{1.5}}
\end{picture}    }
\put(52,5){
\unitlength1.mm
\begin{picture}(8.25,2)
\linethickness{0.15mm}
\put(2,1){\vector(1,0){3.75}}
\put(7.5,1){\line(-1,0){1.75}}
\put(1,1){\circle*{1.5}}
\put(7.5,1){\circle*{1.5}}
\end{picture}  }
\put(0,15.5){
\parbox[t]{15cm}{
\begin{eqnarray}
\label{O9}
&&=\ \ \ \ \ \hspace{3.5cm}=\ {M\over\sqrt{2}}\hspace{5.1cm}
\end{eqnarray}  }}
\end{picture} 

\vspace{7mm}

It is interesting to note that the results (\ref{O5}), (\ref{O7}),
(\ref{O9}) are symmetric under arrow exchange, while those for  
(\ref{O4}), (\ref{O6}), (\ref{O8}) are antisymmetric. This 
consideration can also be extended to the vertices (\ref{V8})-(\ref{V19}),
(\ref{V21})-(\ref{V26}) on the even sublattice $\Lambda_e$, 
when the graphical symbols (\ref{F1})-(\ref{F8}) are also used there. 
For the moment, we denote the different terms in (\ref{V8})-(\ref{V19}),
(\ref{V21})-(\ref{V26}) as subvertices. One finds that the 
subvertices on the even sublattice have exactly the same symmetry 
properties with respect to arrow reversal as their counterparts
on the odd sublattice.\\

\subsection{The calculation of vertex cluster weights}

\parindent1.5em

In order to gain further insight into the structures under 
study it is useful to consider in eqs.\ (\ref{V2})-(\ref{V19}), 
(\ref{V21})-(\ref{V26}) the numerical factor in front of a given 
product of four Grassmann variables (where the sign is fixed by 
the ordering prescription used) as weight of a given subvertex 
(combination of oriented line segments) on the even 
sublattice $\Lambda_e$. 
One then observes that the subvertices (products) in
(\ref{V2})-(\ref{V19}), (\ref{V21})-(\ref{V26}) have exactly 
the same weights as their counterparts on the odd 
sublattice $\Lambda_o$ (cf.\ eqs.\ (\ref{O1})-(\ref{O9})).
Consequently, from now on the distinction 
between even and odd sublattice vertices will no longer 
play any role (except in the discussion of vertex 20
given further below).
Furthermore, one can convince oneself that their weights can easily be 
given by means of a simple rule. 
Each oriented line segment can be associated with a certain weight 
(+1 for straight lines, $\pm 1/\sqrt{2}$ for bent lines). The weight 
for any particular line segment can be read from the upper lines in 
eqs.\ (\ref{S1})-(\ref{S6}).
The weights in vertices 2-19 (eqs.\ (\ref{V2})-(\ref{V19})) 
can then be calculated by simply representing them by two 
oriented line segments (thick lines are understood as
two thin lines with opposite orientations) and then multiplying
the weights of these line segments. For example, this way
one obtains from (\ref{S3}), (\ref{S4}) the negative sign
on the r.h.s.\ of (\ref{V4}), (\ref{V5}). To obtain the weights 
in (\ref{V21})-(\ref{V26}) one simply multiplies the line 
segment weights by another factor of $M$. The same rule applies to 
eqs.\ (\ref{O1})-(\ref{O9}).\\

We are now almost prepared to calculate the weight of any given 
vertex cluster (i.e.\ a connected diagram built of the 
vertices 2-26). One remaining subject to be discussed is
whether our above ordering prescriptions for products of odd sublattice 
Grassmann fields related to the vertices on the even and 
odd sublattices are compatible.
As each cluster can be understood as consisting of 
a certain number of thin line loops we may study the 
ordering problem separately for a single loop (For this
purpose we consider
thick lines as two oppositely oriented thin lines.). 
With an eye to eq.\ (\ref{B4}), one may now easily 
convince oneself that if for the even 
sublattice vertices the ordering prescription applied in 
eqs.\ (\ref{V2})-(\ref{V19}), (\ref{V21})-(\ref{V26})
is used one need not to reorder the Grassmann variables 
related to the odd sublattice points in order to obey
the ordering prescription used in eqs.\ (\ref{O1})-(\ref{O9})
except for one arbitrarily selected point in any single loop.
Consequently, each thin line loop contributes an
additional factor of -1, characteristic of any fermion loop in 
field theory, arising from 
ordering effects. Consequently, thick line loops do not contribute 
any minus sign because they can be understood as two thin line 
loops.\\

After the above discussion it is now straightforward to 
also analyze vertex clusters which contain the vertex 20 
(and its analogue on the odd sublattice)
we omitted so far. The remarkable eq.\ (\ref{R5}) immediately 
suggests how to proceed. Each vertex 20 can be split into 
three different diagrams where the vertex is interpreted
within the loop picture according to the graphical symbols on
the r.h.s.\ of eq.\ (\ref{R5}), and then one may proceed exactly
as for vertex clusters without vertex 20 (In total we have 
($3\times 4 =$) twelve, in part identical, products of Grassmann fields with 
different loop picture interpretations attached to them,
i.e.\ four products for each of the three terms on the r.h.s.\ of 
eq.\ (\ref{R5}).). However, eq.\ (\ref{R5})
stands for a vertex 20 on the even sublattice $\Lambda_e$ and it remains
to be seen whether an analogous interpretation is possible for 
the analogue of vertex 20 on the odd sublattice $\Lambda_o$. The results 
for it read as follows (We here apply the same conventions 
as in eqs.\ (\ref{O1})-(\ref{O9}).).

\parindent0.em
\vspace{7mm}

\unitlength1.mm
\begin{picture}(150,14.5)
\put(3,0){
\unitlength1.mm
\begin{picture}(14.5,14.5)
\linethickness{0.15mm}
\put(0.75,7.25){\vector(1,0){3.75}}
\put(6.25,7.25){\line(-1,0){1.75}}
\put(15.25,7.25){\vector(-1,0){3.75}}
\put(9.75,7.25){\line(1,0){1.75}}
\put(8.75,6.25){\vector(0,-1){3.75}}
\put(8.75,0.75){\line(0,1){1.75}}
\put(7.25,8.25){\vector(0,1){3.75}}
\put(7.25,13.75){\line(0,-1){1.75}}
\put(7.25,7.25){\circle{2}}
\put(7.25,7.25){\circle*{1.5}}
\put(8.75,7.25){\circle{2}}
\put(8.75,7.25){\circle*{1.5}}
\put(15.25,7.25){\circle*{1.5}}
\put(0.75,7.25){\circle*{1.5}}
\put(7.25,13.75){\circle*{1.5}}
\put(8.75,0.75){\circle*{1.5}}
\end{picture} }
\put(33,0){
\unitlength1.mm
\begin{picture}(14.5,14.5)
\linethickness{0.15mm}
\put(9.75,7.25){\vector(1,0){3.75}}
\put(15.25,7.25){\line(-1,0){1.75}}
\put(6.25,7.25){\vector(-1,0){3.75}}
\put(0.75,7.25){\line(1,0){1.75}}
\put(7.25,13.75){\vector(0,-1){3.75}}
\put(7.25,8.25){\line(0,1){1.75}}
\put(8.75,0.75){\vector(0,1){3.75}}
\put(8.75,6.25){\line(0,-1){1.75}}
\put(7.25,7.25){\circle{2}}
\put(7.25,7.25){\circle*{1.5}}
\put(8.75,7.25){\circle{2}}
\put(8.75,7.25){\circle*{1.5}}
\put(15.25,7.25){\circle*{1.5}}
\put(0.75,7.25){\circle*{1.5}}
\put(7.25,13.75){\circle*{1.5}}
\put(8.75,0.75){\circle*{1.5}}
\end{picture}  }
\put(0,15.75){
\parbox[t]{15cm}{
\begin{eqnarray}
\label{O10}
&&=\hspace{9.cm}=\ -1\hspace{2mm}
\end{eqnarray}  }}
\end{picture} 

\vspace{5mm}

\unitlength1.mm
\begin{picture}(150,14.5)
\put(3,0){
\unitlength1.mm
\begin{picture}(14.5,14.5)
\linethickness{0.15mm}
\put(9.75,7.25){\vector(1,0){3.75}}
\put(15.25,7.25){\line(-1,0){1.75}}
\put(0.75,7.25){\vector(1,0){3.75}}
\put(6.25,7.25){\line(-1,0){1.75}}
\put(7.25,8.25){\vector(0,1){3.75}}
\put(7.25,13.75){\line(0,-1){1.75}}
\put(8.75,0.75){\vector(0,1){3.75}}
\put(8.75,6.25){\line(0,-1){1.75}}
\put(7.25,7.25){\circle{2}}
\put(7.25,7.25){\circle*{1.5}}
\put(8.75,7.25){\circle{2}}
\put(8.75,7.25){\circle*{1.5}}
\put(15.25,7.25){\circle*{1.5}}
\put(0.75,7.25){\circle*{1.5}}
\put(7.25,13.75){\circle*{1.5}}
\put(8.75,0.75){\circle*{1.5}}
\end{picture} }
\put(33,0){
\unitlength1.mm
\begin{picture}(14.5,14.5)
\linethickness{0.15mm}
\put(15.25,7.25){\vector(-1,0){3.75}}
\put(9.75,7.25){\line(1,0){1.75}}
\put(6.25,7.25){\vector(-1,0){3.75}}
\put(0.75,7.25){\line(1,0){1.75}}
\put(7.25,13.75){\vector(0,-1){3.75}}
\put(7.25,8.25){\line(0,1){1.75}}
\put(8.75,6.25){\vector(0,-1){3.75}}
\put(8.75,0.75){\line(0,1){1.75}}
\put(7.25,7.25){\circle{2}}
\put(7.25,7.25){\circle*{1.5}}
\put(8.75,7.25){\circle{2}}
\put(8.75,7.25){\circle*{1.5}}
\put(15.25,7.25){\circle*{1.5}}
\put(0.75,7.25){\circle*{1.5}}
\put(7.25,13.75){\circle*{1.5}}
\put(8.75,0.75){\circle*{1.5}}
\end{picture} }
\put(63,0){
\unitlength1.mm
\begin{picture}(14.5,14.5)
\linethickness{0.15mm}
\put(15.25,7.25){\vector(-1,0){3.75}}
\put(9.75,7.25){\line(1,0){1.75}}
\put(6.25,7.25){\vector(-1,0){3.75}}
\put(0.75,7.25){\line(1,0){1.75}}
\put(8.75,8.25){\vector(0,1){3.75}}
\put(8.75,13.75){\line(0,-1){1.75}}
\put(7.25,0.75){\vector(0,1){3.75}}
\put(7.25,6.25){\line(0,-1){1.75}}
\put(7.25,7.25){\circle{2}}
\put(7.25,7.25){\circle*{1.5}}
\put(8.75,7.25){\circle{2}}
\put(8.75,7.25){\circle*{1.5}}
\put(15.25,7.25){\circle*{1.5}}
\put(0.75,7.25){\circle*{1.5}}
\put(8.75,13.75){\circle*{1.5}}
\put(7.25,0.75){\circle*{1.5}}
\end{picture}  }
\put(93,0){
\unitlength1.mm
\begin{picture}(14.5,14.5)
\linethickness{0.15mm}
\put(9.75,7.25){\vector(1,0){3.75}}
\put(15.25,7.25){\line(-1,0){1.75}}
\put(0.75,7.25){\vector(1,0){3.75}}
\put(6.25,7.25){\line(-1,0){1.75}}
\put(8.75,13.75){\vector(0,-1){3.75}}
\put(8.75,8.25){\line(0,1){1.75}}
\put(7.25,6.25){\vector(0,-1){3.75}}
\put(7.25,0.75){\line(0,1){1.75}}
\put(7.25,7.25){\circle{2}}
\put(7.25,7.25){\circle*{1.5}}
\put(8.75,7.25){\circle{2}}
\put(8.75,7.25){\circle*{1.5}}
\put(15.25,7.25){\circle*{1.5}}
\put(0.75,7.25){\circle*{1.5}}
\put(8.75,13.75){\circle*{1.5}}
\put(7.25,0.75){\circle*{1.5}}
\end{picture} }
\put(0,16.25){
\parbox[t]{15cm}{
\begin{eqnarray}
\label{O11}
&&=\hspace{2.7cm}=\hspace{2.7cm}=\hspace{3cm}=\ - {1\over 2}\hspace{2mm}
\end{eqnarray} }}
\end{picture} 

\vspace{5mm}

In the above equations for ordering purposes we have 
applied a certain interpretation in terms of oriented 
line segments. However, each of the above six pictures allows 
two different interpretations in terms of oriented line 
segments (leading in total again to twelve different graphical terms). 
Consequently, one may decide to apply one ordering 
in a certain fraction $\alpha\in {\rm{\bf R}}$ 
of cases (i.e.\ for those cases shown in (\ref{O10}), (\ref{O11})) 
and in another fraction $(1 -\alpha)$ the other, 
alternative prescription. The weight for each subvertex then 
has to be multiplied by $\alpha$ or $(1 -\alpha)$
respectively in order to obtain the correct final result 
for the partition function.
The two vertices in eq.\ (\ref{O10}) can both 
be interpreted in an alternative way (relative to the 
one shown in eq.\ (\ref{O10})) as two oriented bent line 
segments passing each other without intersection. 
The four vertices in eqs.\ (\ref{O11})
each allow an interpretation as either two oriented bent line segments
passing each other without intersection (as shown) 
or as two oriented straight line segments with an intersection. 
In order to make convenient contact with our above discussion for 
clusters without the vertex 20 (and its analogue on the odd
sublattice), it now turns out that for eq.\ (\ref{O10}) 
the choice $\alpha = 1/2$ is to be made while for eq.\ (\ref{O11})
$\alpha = -1$ has to be selected\footnote{Clearly, the final 
result for the partition function does not depend on the 
choice of $\alpha$; however the present choice admits a particularly
simple way of describing contributions to the partition 
function.}. If one does so the following 
simple rule for treating the vertex 20 (and its analogue on the
odd sublattice) emerges. Replace in any vertex cluster under 
consideration each vertex 20 (and its analogue 
on the odd sublattice) by two line segments 
according to the graphical rule

\vspace{5mm}

\unitlength1.mm
\begin{picture}(150,15)
\put(10,0){
\unitlength1.mm
\begin{picture}(15,15)
\linethickness{0.15mm}
\put(2,7.5){\line(1,0){11}}
\put(7.5,2){\line(0,1){11}}
\end{picture} }
\put(50,0){
\unitlength1.mm
\begin{picture}(15,15)
\linethickness{0.15mm}
\put(2,7.5){\line(1,0){3.5}}
\put(9.5,7.5){\line(1,0){3.5}}
\put(5.5,9.5){\oval(4,4)[br]}
\put(9.5,5.5){\oval(4,4)[tl]}
\put(7.5,2){\line(0,1){3.5}}
\put(7.5,9.5){\line(0,1){3.5}}
\end{picture} }
\put(80,0){
\unitlength1.mm
\begin{picture}(15,15)
\linethickness{0.15mm}
\put(2,7.5){\line(1,0){3.5}}
\put(9.5,7.5){\line(1,0){3.5}}
\put(9.5,9.5){\oval(4,4)[bl]}
\put(5.5,5.5){\oval(4,4)[tr]}
\put(7.5,2){\line(0,1){3.5}}
\put(7.5,9.5){\line(0,1){3.5}}
\end{picture} }
\put(110,0){
\unitlength1.mm
\begin{picture}(15,15)
\linethickness{0.15mm}
\put(2,7.5){\line(1,0){10.5}}
\put(7.5,7.5){\oval(4,4)[bl]}
\put(7.5,7.5){\oval(4,4)[tl]}
\put(7.5,2){\line(0,1){3.5}}
\put(7.5,9.5){\line(0,1){3.5}}
\end{picture} }
\put(0,16){
\parbox[t]{15cm}{
\begin{eqnarray}
\label{rule}
&&\longrightarrow\hspace{2.6cm}+\hspace{2.6cm}+\hspace{2.cm}
\ \ \ \ .\hspace{-1.cm}
\end{eqnarray}  }}
\end{picture}

\vspace{5mm}

Each of the line segments can be oriented, consequently each 
of the three terms on the r.h.s.\ of eq.\ (\ref{rule}) yields
four different diagrams of two oriented line segments (twelve
different terms in total). While 
clusters without the vertex 20 (and its analogue on the odd
sublattice) contain self-avoiding ($k_l = 1$) thin line loops only, the 
last term on the r.h.s.\ of eq.\ (\ref{rule}) now allows
($k_l = 1$) thin line loops with intersections also. The weight of any
of the twelve possible diagrams containing now instead of the vertex 20
two oriented ($k_l = 1$)
thin line segments, passing each other without intersection 
or intersecting each other, can simply be calculated by 
taking separately the weights of each oriented line segment from 
eqs.\ (\ref{S1})-(\ref{S6}) and multiplying them.
This concludes the discussion of the vertex 20.\\

\parindent1.5em

The rule for calculating the weight of 
a vertex cluster consequently is the following. 
Replace any vertex 20 (and its analogue on the odd 
sublattice) according to the rule (\ref{rule}) by
two ($k_l = 1$) thin line segments (If $P_{20}$ is the number
of its occurrences in a given vertex cluster, this
rule leads to $3^{\displaystyle P_{20}}$ different configurations.
The latter term from now on always denotes the diagrams
generated from a vertex cluster by application of the 
rule (\ref{rule}), or (\ref{colourrule}).).
Provide any of the, say, $m$ ($k_l = 1$) thin line loops (without 
reinterpreting thick lines as two parallel thin lines) 
with an orientation. This leads for a given configuration
to $2^m$ different graphs (From now on the latter term denotes
diagrams generated this way from configurations.).
Represent the obtained graphs completely as a collection 
of, say, $n$ oriented thin line loops (with thick lines 
interpreted as two oppositely oriented thin lines). 
In total, this leads for any original vertex cluster
under consideration to different $3^{\displaystyle P_{20}}\ 2^m$ 
graphs with a certain number of oriented thin line loops each
(this number may differ from graph to graph).
Calculate the weight of each 
single oriented thin loop of length $l$ by multiplying 
the weights of the $l$ combinations of oriented line segments 
it is built of and attach the ordering factor -1. 
The weight of each graph is given by the product of 
the $n$ oriented thin line loops it can be represented by.
Any vertex cluster consequently has as weight the sum
of the weights of the $3^{\displaystyle P_{20}}\ 2^m$ graphs
generated from it.\\

In order to make further progress, first we have to study the 
weight of an arbitrary oriented thin line loop $L$. According 
to eqs.\ (\ref{S1})-(\ref{S6}) the absolute value of the 
weight is given by the number of corners $C(L)$ of the loop 
as well as the number of occurrences of the vertices 21-26, 
denoted by $P_M$, and reads 
$2^{\displaystyle -C(L)/2}\ M^{\displaystyle P_M}$. According to
the above considerations the sign of the loop weight is given by 
$(-1)^{\displaystyle (C_{-}(L) + 1)}$, 
where $C_{-}(L)$ is the number of (oriented)
right-up, up-right segments (cf.\ the second term on the 
r.h.s.\ of eqs.\ (\ref{S3}), (\ref{S4})). 
Now, it is useful to take into account the equation 
\begin{eqnarray}
\label{B10a}
C_{-}(L)&\equiv& q(L) + 1\ ({\rm mod}\ 2)\ \  ,
\end{eqnarray} 
where $q(L)$ is the number of self-intersections of the loop $L$.
A proof of eq.\ (\ref{B10a}) can be found in Appendix A. From 
eq.\ (\ref{B10a}) immediately follows that the weight of a
thin line loop under consideration 
is given by\footnote{Incidentally, $P_M$ is always even.
This can easily be seen for any given vertex cluster by
simply dropping all thick lines from it, i.e.\ by replacing
all vertices 8-19 by vertices 21-26.}
\begin{eqnarray}
\label{B12}
&&(-1)^{\displaystyle q(L)}\ 
2^{\displaystyle -C(L)/2}\ M^{\displaystyle P_M}\ . 
\end{eqnarray}
This formula immediately also applies
if $L$ is understood to be a graph of, say, $n$ thin line loops
($q(L)$ can still be understood as the number of intersections;
however only self-intersections of loops are really important because
the number of intersections of different loops is always even.).
This graph of loops can exhibit $2^m$ different combinations
of orientations if the configuration (obtained from a vertex 
cluster under consideration by application of the rule 
(\ref{rule})) it is generated from
contains $m$ ($k_l = 1$) thin line loops. However, $q(L)$ may differ for 
these cases. Consequently, cancellations may and will occur among the 
$3^{\displaystyle P_{20}}\ 2^m$ graphs generated from
a given vertex cluster. In the following subsection we continue 
with some observations in this respect.\\

\subsection{From oriented to coloured loops}

\parindent1.5em

For the moment, in order to simplify the discussion let us 
only consider vertex clusters without any 
vertex 20. We show that any vertex cluster $L$ under consideration  
containing a ($k_l = 1$) thin line loop with an odd number 
of the vertices 8-19 has vanishing weight. Let $P_T$ be the number
of the vertices 8-19 in a certain  ($k_l = 1$) thin line loop $A$ in $L$. 
Now, generate from any configuration with 
$P_T(A) \equiv 1\ ({\rm mod}\ 2)$ according to the procedure
explained above $2^m$ 
graphs of oriented thin line loops (with thick lines interpreted 
as two oppositely oriented thin lines). There are, say, $P_{T1}$ 
vertices 8-19 in the ($k_l = 1$) thin line loop 
$A$ whose attached thick lines
start and end at the loop $A$, consequently 
$P_{T1}\equiv 0\ ({\rm mod}\ 2)$. If these thick lines are 
interpreted as thin lines, the number of thin line intersections 
is zero (modulo 2). There are $P_{T2}$ 
vertices 8-19 in the ($k_l = 1$) thin line 
loop $A$ whose attached thick lines start
at the loop $A$ and end at some other ($k_l = 1$) thin line loop. 
If these thick lines are 
interpreted as thin lines, the number of thick lines starting from 
the loop $A$ and containing one or zero (modulo 2) thin line intersections 
are $P_{T2o}$ and $P_{T2e}$ respectively. It holds
\begin{eqnarray}
\label{B13}
P_T(A)\ = \ P_{T1}\ +\ P_{T2o}\ +\ P_{T2e}&\equiv& 1\ ({\rm mod}\ 2)
\ \ \ .
\end{eqnarray}
Consequently, either $P_{T2o}$ or $P_{T2e}$ is odd
(and the other is even, of course). If one now 
reverses the orientation of just the ($k_l = 1$) 
thin line loop $A$ (with all other 
orientations fixed) the numbers $P_{T2o}$ and $P_{T2e}$
interchange (and 
$\Delta q = \Delta P_{T2o} \equiv 1\ ({\rm mod}\ 2)$ for the 
graphs with the opposite orientation of $A$). From 
eq.\ (\ref{B12}) one immediately 
recognizes that after summation over the two opposite 
orientations of the loop $A$ the weight of any configuration
containing $A$ vanishes.\\

We now draw some first conclusions from the insight just obtained. 
We demonstrate that any vertex cluster $L$ can be built by the 
overlap of two sets of self-avoiding loops (Self-avoidance 
here means that also intersections of different loops within a given 
set are not allowed.). 
We distinguish the loops out of the two different
sets by attaching a colour, say red and blue, to each sort 
of loop. We proceed as follows. Divide any given ($k_l = 1$) thin line 
loop of a vertex cluster $L$ into sections 
of alternating colour where the colour 
changes at any vertex 8-19 (This is possible, because any 
($k_l = 1$) thin line loop contributing to the partition function 
contains an even number of vertices 8-19, as we have shown in 
the preceding paragraph.). This procedure yields two 
different colourings for each ($k_l = 1$) thin line loop in $L$ (They just
differ by an interchange between red and blue.). These two 
colourings stand in correspondence to the two different 
orientations of any ($k_l = 1$) thin line loop in $L$. We can then 
interpret any thick line in $L$ as two parallel thin lines 
of different colour. It is clear that the described 
procedure yields a covering of the given vertex cluster $L$
by two sets of self-avoiding, differently coloured 
loops (at each vertex meet, at most, two links of the same
colour only).
Consequently, in view of eq.\ (\ref{B12}), the weight of any 
vertex cluster (not containing any vertex 20) which is not 
zero must be positive. The graphs generated from vertex 
clusters with positive weight can be thought of as being 
constructed from the overlap of two differently coloured 
sets of self-avoiding loops (with monomer weight
$z = M$ and bending rigidity $\eta = 1/\sqrt{2}$).
Their weight is given by the product of the weights of the coloured 
loops they are built of. The number of different graphs
corresponding to a vertex cluster is exactly given by the number
of different coverings of it by means of two sets of 
differently coloured self-avoiding loops.\\

We would now like to extend the discussion to vertex clusters
containing vertices of type no.\ 20. The best strategy to do so
seems to be to transform any vertex cluster containing vertices 
of type no.\ 20 into ``equivalent'' vertex clusters without these
(The problem we are interested in is primarily a topological one.).
By ``equivalent'' we mean vertex clusters which have the 
same weight as the original one up to some positive constant
which can easily be taken care of. The idea is to ``stretch''
any vertex 20 into two parts in such a way that in the 
``equivalent'' cluster both parts are connected by a thick
line. For any given column (row) of lattice points which 
contains a vertex 20 the ``stretching'' can be 
facilitated by inserting two additional lattice point columns 
(rows) to the left (top) and right (bottom) of the column (row)
under consideration (The procedure is to be performed for 
each individual vertex 20 separately.). Pictorially, the procedure can 
be represented by the rule (for columns)

\vspace{5mm}
\noindent
\unitlength1.mm
\begin{picture}(150,15)
\put(10,0){
\unitlength1.mm
\begin{picture}(15,15)
\linethickness{0.15mm}
\put(2,7.5){\line(1,0){11}}
\put(7.5,2){\line(0,1){11}}
\end{picture} }
\put(50,0){
\unitlength1.mm
\begin{picture}(15,15)
\linethickness{0.15mm}
\put(2,7.5){\line(1,0){5.5}}
\put(18.5,7.5){\line(1,0){5.5}}
\put(18.5,2){\line(0,1){5.5}}
\put(7.5,7.5){\line(0,1){5.5}}
\linethickness{0.6mm}
\put(7.5,7.5){\line(1,0){11}}
\end{picture} }
\put(91,0){
\unitlength1.mm
\begin{picture}(15,15)
\linethickness{0.15mm}
\put(2,7.5){\line(1,0){5.5}}
\put(18.5,7.5){\line(1,0){5.5}}
\put(7.5,2){\line(0,1){5.5}}
\put(18.5,7.5){\line(0,1){5.5}}
\linethickness{0.6mm}
\put(7.5,7.5){\line(1,0){11}}
\end{picture} }
\put(0,16){
\parbox[t]{15cm}{
\begin{eqnarray}
\label{colourrule}
&&\longrightarrow\hspace{3.7cm}+\hspace{3.5cm}.
\end{eqnarray}  }}
\end{picture}

\vspace{5mm}

\noindent
Any other vertex 2-26 in the given column can also easily be 
``stretched'' (shifted to the left or right)
by appropriately inserting further horizontal thick or 
($k_l = 1$) thin lines 
(and filling the remaining sites emerging in the ``stretching''
procedure by monomers). Although one might be tempted to do so, 
the rule (\ref{colourrule}) should not be 
confused with the rule (\ref{rule}) which has a different 
algebraic meaning in terms of combinations of Grassmann variables.
It is however a way of dealing with the vertex 20 which is 
alternative to (\ref{rule}).
Let us be more specific to illustrate (\ref{colourrule}).
For example, for a vertex 20 on the even sublattice $\Lambda_e$
one finds by means of 
the above procedure the expression ($x$ denotes 
the coordinate of the vertex 20 on the ``unstretched'' lattice,
which on the ``stretched'' lattice becomes the middle point 
of the introduced thick lines.
On the ``stretched'' lattice the Grassmann integrations at the 
points $x - e_1$, $x$, $x + e_1$ have been carried out.)

\begin{eqnarray}
\label{B14}
&&-\ \chi_1(x + e_1 - e_2)\ \bar\psi_1(x + 2 e_1)
\ \chi_2(x - e_1 + e_2)\ \bar\psi_2(x - 2 e_1)/2\nonumber\\[0.3cm]
&&-\ \psi_1(x - 2 e_1)\ \bar\chi_1(x - e_1 + e_2)
\ \chi_1(x + e_1 - e_2)\ \bar\psi_1(x + 2 e_1)/2\nonumber\\[0.3cm]
&&-\ \psi_2(x + 2 e_1)\ \bar\chi_2(x + e_1 - e_2)
\ \chi_2(x - e_1 + e_2)\ \bar\psi_2(x - 2 e_1)/2\nonumber\\[0.3cm]
&&-\ \psi_1(x - 2 e_1)\ \bar\chi_1(x - e_1 + e_2)
\ \psi_2(x + 2 e_1)\ \bar\chi_2(x + e_1 - e_2)/2\nonumber\\[0.3cm]
&&-\ \chi_1(x - e_1 - e_2)\ \bar\psi_1(x + 2 e_1)
\ \chi_2(x + e_1 + e_2)\ \bar\psi_2(x - 2 e_1)/2\nonumber\\[0.3cm]
&&-\ \psi_2(x + 2 e_2)\ \bar\chi_1(x + e_1 + e_2)
\ \chi_1(x - e_1 - e_2)\ \bar\psi_2(x - 2 e_1)/2\nonumber\\[0.3cm]
&&-\ \psi_1(x - 2 e_1)\ \bar\chi_2(x - e_1 - e_2)
\ \chi_2(x + e_1 + e_2)\ \bar\psi_1(x + 2 e_2)/2\nonumber\\[0.3cm]
&&-\ \psi_1(x - 2 e_1)\ \bar\chi_1(x + e_1 + e_2)
\ \psi_2(x + 2 e_1)\ \bar\chi_2(x - e_1 - e_2)/2\ \ .
\end{eqnarray}  

\noindent
The first four terms correspond to the first term on the r.h.s.\ 
of (\ref{colourrule}) while the last four terms correspond  
to the second term on the r.h.s.\ of (\ref{colourrule}).
If one shifts back each argument in (\ref{B14}) by either plus or
minus $e_1$ one reobtains the expression on the r.h.s.\ of 
eq.\ (\ref{V20}). Let us also discuss the rule (\ref{colourrule}) from 
the point of view of a vertex 20 on the odd sublattice $\Lambda_o$.
Some of the corresponding results read as follows (Again, the Grassmann 
integration has been performed at the points $x - e_1$, $x$, 
$x + e_1$, where $x$ denotes the coordinates of the vertex
20 on the odd sublattice.).

\vspace{7mm}

\noindent
\unitlength1.mm
\begin{picture}(150,14.5)
\put(3,0){
\unitlength1.mm
\begin{picture}(14.5,14.5)
\linethickness{0.15mm}
\put(15.5,7.25){\vector(1,0){3.75}}
\put(21,7.25){\line(-1,0){1.75}}
\put(8.5,7.25){\vector(1,0){3.75}}
\put(14,7.25){\line(-1,0){1.75}}
\put(1.5,7.25){\vector(1,0){3.75}}
\put(7,7.25){\line(-1,0){1.75}}
\put(21.75,8.){\vector(0,1){3.75}}
\put(21.75,13.5){\line(0,-1){1.75}}
\put(21.75,7.25){\circle{2}}
\put(21.75,7.25){\circle*{1.5}}
\put(7.75,7.25){\circle{2}}
\put(7.75,7.25){\circle*{1.5}}
\put(0.75,7.25){\circle*{1.5}}
\put(14.75,7.25){\circle*{1.5}}
\put(21.75,14.25){\circle*{1.5}}

\put(14,5.75){\vector(-1,0){3.75}}
\put(8.5,5.75){\line(1,0){1.75}}
\put(21,5.75){\vector(-1,0){3.75}}
\put(15.5,5.75){\line(1,0){1.75}}
\put(28,5.75){\vector(-1,0){3.75}}
\put(22.5,5.75){\line(1,0){1.75}}
\put(7.75,5){\vector(0,-1){3.75}}
\put(7.75,-0.5){\line(0,1){1.75}}
\put(21.75,5.75){\circle{2}}
\put(21.75,5.75){\circle*{1.5}}
\put(28.75,5.75){\circle*{1.5}}
\put(7.75,5.75){\circle{2}}
\put(7.75,5.75){\circle*{1.5}}
\put(14.75,5.75){\circle*{1.5}}
\put(7.75,-1.25){\circle*{1.5}}

\end{picture} }
\put(46,0){
\unitlength1.mm
\begin{picture}(14.5,14.5)
\linethickness{0.15mm}
\put(21,7.25){\vector(-1,0){3.75}}
\put(15.5,7.25){\line(1,0){1.75}}
\put(14,7.25){\vector(-1,0){3.75}}
\put(8.5,7.25){\line(1,0){1.75}}
\put(7,7.25){\vector(-1,0){3.75}}
\put(1.5,7.25){\line(1,0){1.75}}
\put(21.75,13.5){\vector(0,-1){3.75}}
\put(21.75,8){\line(0,1){1.75}}
\put(21.75,7.25){\circle{2}}
\put(21.75,7.25){\circle*{1.5}}
\put(7.75,7.25){\circle{2}}
\put(7.75,7.25){\circle*{1.5}}
\put(0.75,7.25){\circle*{1.5}}
\put(14.75,7.25){\circle*{1.5}}
\put(21.75,14.25){\circle*{1.5}}

\put(8.5,5.75){\vector(1,0){3.75}}
\put(14,5.75){\line(-1,0){1.75}}
\put(15.5,5.75){\vector(1,0){3.75}}
\put(21,5.75){\line(-1,0){1.75}}
\put(22.5,5.75){\vector(1,0){3.75}}
\put(28,5.75){\line(-1,0){1.75}}
\put(7.75,-0.75){\vector(0,1){3.75}}
\put(7.75,4.75){\line(0,-1){1.75}}
\put(21.75,5.75){\circle{2}}
\put(21.75,5.75){\circle*{1.5}}
\put(28.75,5.75){\circle*{1.5}}
\put(7.75,5.75){\circle{2}}
\put(7.75,5.75){\circle*{1.5}}
\put(14.75,5.75){\circle*{1.5}}
\put(7.75,-1.25){\circle*{1.5}}
\end{picture}  }
\put(0,15.75){
\parbox[t]{15cm}{
\begin{eqnarray}
\label{O12}
&&=\hspace{4.cm}=\ -{1\over 2}\hspace{2.5cm}
\end{eqnarray}  }}
\end{picture} 

\vspace{5mm}

\noindent
\unitlength1.mm
\begin{picture}(150,14.5)
\put(3,0){
\unitlength1.mm
\begin{picture}(14.5,14.5)
\linethickness{0.15mm}
\put(28,7.25){\vector(-1,0){3.75}}
\put(22.5,7.25){\line(1,0){1.75}}
\put(21,7.25){\vector(-1,0){3.75}}
\put(15.5,7.25){\line(1,0){1.75}}
\put(14,7.25){\vector(-1,0){3.75}}
\put(8.5,7.25){\line(1,0){1.75}}
\put(7,7.25){\vector(-1,0){3.75}}
\put(1.5,7.25){\line(1,0){1.75}}
\put(21.75,7.25){\circle{2}}
\put(21.75,7.25){\circle*{1.5}}
\put(7.75,7.25){\circle{2}}
\put(7.75,7.25){\circle*{1.5}}
\put(0.75,7.25){\circle*{1.5}}
\put(14.75,7.25){\circle*{1.5}}
\put(28.75,7.25){\circle*{1.5}}

\put(7,5.75){\vector(1,0){3.75}}
\put(12.5,5.75){\line(-1,0){1.75}}
\put(14,5.75){\vector(1,0){3.75}}
\put(19.5,5.75){\line(-1,0){1.75}}
\put(6.25,-0.75){\vector(0,1){3.75}}
\put(6.25,4.75){\line(0,-1){1.75}}
\put(20.25,6.5){\vector(0,1){3.75}}
\put(20.25,12){\line(0,-1){1.75}}
\put(20.25,5.75){\circle{2}}
\put(20.25,5.75){\circle*{1.5}}
\put(6.25,5.75){\circle{2}}
\put(6.25,5.75){\circle*{1.5}}
\put(13.25,5.75){\circle*{1.5}}
\put(6.25,-1.25){\circle*{1.5}}
\put(20.25,12.75){\circle*{1.5}}
\end{picture}  }

\put(46,0){
\unitlength1.mm
\begin{picture}(14.5,14.5)
\linethickness{0.15mm}
\put(22.5,7.25){\vector(1,0){3.75}}
\put(28,7.25){\line(-1,0){1.75}}
\put(15.5,7.25){\vector(1,0){3.75}}
\put(21,7.25){\line(-1,0){1.75}}
\put(8.5,7.25){\vector(1,0){3.75}}
\put(14,7.25){\line(-1,0){1.75}}
\put(1.5,7.25){\vector(1,0){3.75}}
\put(7,7.25){\line(-1,0){1.75}}
\put(21.75,7.25){\circle{2}}
\put(21.75,7.25){\circle*{1.5}}
\put(7.75,7.25){\circle{2}}
\put(7.75,7.25){\circle*{1.5}}
\put(0.75,7.25){\circle*{1.5}}
\put(14.75,7.25){\circle*{1.5}}
\put(28.75,7.25){\circle*{1.5}}

\put(12.5,5.75){\vector(-1,0){3.75}}
\put(7,5.75){\line(1,0){1.75}}
\put(19.5,5.75){\vector(-1,0){3.75}}
\put(14,5.75){\line(1,0){1.75}}
\put(6.25,5){\vector(0,-1){3.75}}
\put(6.25,-0.5){\line(0,1){1.75}}
\put(20.25,12){\vector(0,-1){3.75}}
\put(20.25,6.5){\line(0,1){1.75}}
\put(20.25,5.75){\circle{2}}
\put(20.25,5.75){\circle*{1.5}}
\put(6.25,5.75){\circle{2}}
\put(6.25,5.75){\circle*{1.5}}
\put(13.25,5.75){\circle*{1.5}}
\put(6.25,-1.25){\circle*{1.5}}
\put(20.25,12.75){\circle*{1.5}}
\end{picture}  }
\put(0,15.75){
\parbox[t]{15cm}{
\begin{eqnarray}
\label{O13}
&&=\hspace{4.cm}=\ \ \ \; {1\over 2}\hspace{2.5cm}
\end{eqnarray}  }}
\end{picture} 

\vspace{5mm}

\noindent
In these equations we again apply the graphical conventions 
as explained previously (cf.\ eqs.\ (\ref{S1})-(\ref{S6}),
(\ref{F9})).
The equations (\ref{O12}), (\ref{O13}) relate to the first
term on the r.h.s.\ of (\ref{colourrule}), analogous results 
can be given for the second term. Eq.\ (\ref{O12}) corresponds
to eq.\ (\ref{O10}) (Remember that for the ordering 
choice selected in eq.\ (\ref{O10}) we have applied the weight
$\alpha = 1/2$, consequently we find $-1/2$ on the r.h.s.\
of eq.\ (\ref{O12}).). Eq.\ (\ref{O13}) corresponds
to the last two terms in eq.\ (\ref{O11}) (However,
this time for the ordering 
choice selected in eq.\ (\ref{O11}) we have applied the weight
$\alpha = 0$. Consequently, eq.\ (\ref{O11}) is exclusively 
interpreted as intersection of oriented straight line 
segments. The corresponding reordering of the Grassmann
variables introduces a relative minus sign.
Therefore, we find $1/2$ on the r.h.s.\ of eq.\ (\ref{O13}).).\\

The rule (\ref{colourrule}) now allows to efficiently 
handle the colouring problem for vertex clusters $L$
containing vertex 20 and also to calculate their
weights. For any given vertex cluster the rule (\ref{colourrule}) 
introduces $2^{\displaystyle P_{20}}$ different 
configurations free of any vertex 20. As discussed 
above the summation over orientations of ($k_l = 1$) thin line 
loops for these configurations (i.e.\ the summation over graphs)
lets only those survive which have 
an even number of vertices 8-19 in any ($k_l = 1$) thin line loop.
The sum of the weights of these configurations with
non-vanishing (and consequently positive) weights
yields the total weight of any vertex cluster containing
vertices of type no.\ 20. For any graph generated from 
a configuration with nonvanishing weight the 
described colouring procedure
can be carried out and then the rule (\ref{colourrule}) 
can be reversed without any problem (It depends on the 
shape of the configuration whether both terms on the r.h.s.\
of (\ref{colourrule}) lead to diagrams with non-vanishing
weight.).
Consequently, the above insight for the 
calculation of the weight of vertex clusters without 
any vertex 20 immediately carries over to any 
configuration generated from an arbitrary vertex cluster.
Finally, it is clear that any overlap of two sets
of self-avoiding loops generates a vertex cluster
constructed out of the vertices 2-26. The two possible 
orientations of ($k_l = 1$) thin line loops correspond to 
the colour exchange among the two sets (or, more 
precisely, to the exchange of the loop configurations
among the two sets of different colour).\\

\subsection{The two-colour loop model}

We arrive at the following picture for the one-flavour 
Thirring model with Wilson fermions. Any diagram
(and its weight)
contributing to the partition function $Z_\Lambda$
can be regarded as being generated by the overlap 
of two differently coloured sets of self-avoiding loops
with monomer weight $z = M$ and bending rigidity 
$\eta = 1/\sqrt{2}$. Only the monomer 
weight $M^2 - 2G$ (eq.\ (\ref{V1})) of the Thirring model
differs from the product of the monomer weights of the 
two differently coloured self-avoiding loop models $M^2$.
However, for free fermions ($G = 0$) there is complete 
agreement. Consequently, the structure of the 
statistical model equivalent to the Thirring model with 
Wilson fermions is very close to that for a free Dirac
fermion. Therefore, for $G = 0$ we find
\begin{eqnarray}
\label{B15}
Z_\Lambda&=&\left( Z_\Lambda\left[M,{1\over\sqrt{2}}\right]\right)^2\ \ ,
\end{eqnarray}
where $Z_\Lambda[z,\eta]$ is the partition function for
the self-avoiding loop model with monomer weight $z$
and bending rigidity $\eta$
as defined in \cite{scharn2} (see also further below). 
This result is not astonishing
because the self-avoiding loop model with bending 
rigidity $\eta = 1/\sqrt{2}$ is known \cite{pri1} (see also
\cite{pri2}), \cite{blu} to be a free fermion model
\cite{fan1}, \cite{fan2} (cf.\ also sect.\ 4 of \cite{scharn2}).
The model has second order phase transitions (within the
Ising model universality class) at $z = M = 2$ (this 
corresponds to $\kappa = 1/4$) and 
$z = M = 0$ (where it is equivalent to a 6-vertex model).
At $z = M = 0$ the central charge $c$ of the self-avoiding 
loop model with bending rigidity $\eta = 1/\sqrt{2}$ is 
$c = 1$ \cite{kar4}, \cite{kar5} while, as has been 
argued in \cite{scharn2}, at $z = M = 2$ it should have 
$c = 1/2$ which is rather obvious from eq.\ (\ref{B15}) 
because any free Dirac fermion has $c = 1$.\\

Let us now specify in detail the statistical model 
equivalent to the one-flavour lattice Thirring model with
Wilson fermions (and Wilson parameter $r = 1$). This 
statistical model is a two-color loop model
with bending rigidity $\eta = 1/\sqrt{2}$ and monomer
weight $M^2 - 2G$. The vertices and weights of this
model can best be described as follows. Consider the 
(one-colour) self-avoiding loop model with monomer weight $z = M$ and
bending rigidity $\eta = 1/\sqrt{2}$ (For the vertices see
Fig.\ 1.). The weights are
\begin{eqnarray}
\label{B16}
\omega_1 &=& z\ =\ M\ \ \ \ ,\\[0.3cm]
\omega_3 &=& \omega_4\ =\ 1\ \ \ \ ,\\[0.3cm]
\omega_5&=& \omega_6\ =\ \omega_7\ =\ \omega_8\ =\ 
\eta\ =\ {1\over\sqrt{2}}\ \ \ \ .
\end{eqnarray}

\begin{table}[t]
\unitlength1.mm
\begin{picture}(150,40)
\put(0,10){
\unitlength1.mm
\begin{picture}(15,15)
\linethickness{0.15mm}
\put(12,7.5){\line(1,0){1}}
\put(10,7.5){\line(1,0){1}}
\put( 8,7.5){\line(1,0){1}}
\put( 6,7.5){\line(1,0){1}}
\put( 4,7.5){\line(1,0){1}}
\put( 2,7.5){\line(1,0){1}}
\put(7.5,12){\line(0,1){1}}
\put(7.5,10){\line(0,1){1}}
\put(7.5, 8){\line(0,1){1}}
\put(7.5, 6){\line(0,1){1}}
\put(7.5, 4){\line(0,1){1}}
\put(7.5, 2){\line(0,1){1}}
\end{picture}  }
\put(18,10){
\unitlength1.mm
\begin{picture}(15,15)
\linethickness{0.15mm}
\put(12,7.5){\line(1,0){1}}
\put(10,7.5){\line(1,0){1}}
\put( 8,7.5){\line(1,0){1}}
\put( 6,7.5){\line(1,0){1}}
\put( 4,7.5){\line(1,0){1}}
\put( 2,7.5){\line(1,0){1}}
\put(7.5,2){\line(0,1){11}}
\end{picture}  }
\put(36,10){
\unitlength1.mm
\begin{picture}(15,15)
\linethickness{0.15mm}
\put(7.5,12){\line(0,1){1}}
\put(7.5,10){\line(0,1){1}}
\put(7.5, 8){\line(0,1){1}}
\put(7.5, 6){\line(0,1){1}}
\put(7.5, 4){\line(0,1){1}}
\put(7.5, 2){\line(0,1){1}}
\put(2,7.5){\line(1,0){11}}
\end{picture}  }
\put(54,10){
\unitlength1.mm
\begin{picture}(15,15)
\linethickness{0.15mm}
\put( 6,7.5){\line(1,0){1}}
\put( 4,7.5){\line(1,0){1}}
\put( 2,7.5){\line(1,0){1}}
\put(7.5,12){\line(0,1){1}}
\put(7.5,10){\line(0,1){1}}
\put(7.5, 8){\line(0,1){1}}
\put(7.2,7.5){\line(1,0){5.8}}
\put(7.5,2){\line(0,1){5.8}}
\end{picture}  }
\put(72,10){
\unitlength1.mm
\begin{picture}(15,15)
\linethickness{0.15mm}
\put(12,7.5){\line(1,0){1}}
\put(10,7.5){\line(1,0){1}}
\put( 8,7.5){\line(1,0){1}}
\put(7.5, 6){\line(0,1){1}}
\put(7.5, 4){\line(0,1){1}}
\put(7.5, 2){\line(0,1){1}}
\put(2,7.5){\line(1,0){5.8}}
\put(7.5,7.2){\line(0,1){5.8}}
\end{picture}  }
\put(90,10){
\unitlength1.mm
\begin{picture}(15,15)
\linethickness{0.15mm}
\put( 6,7.5){\line(1,0){1}}
\put( 4,7.5){\line(1,0){1}}
\put( 2,7.5){\line(1,0){1}}
\put(7.5, 6){\line(0,1){1}}
\put(7.5, 4){\line(0,1){1}}
\put(7.5, 2){\line(0,1){1}}
\put(7.2,7.5){\line(1,0){5.8}}
\put(7.5,7.2){\line(0,1){5.8}}
\end{picture}  }
\put(108,10){
\unitlength1.mm
\begin{picture}(15,15)
\linethickness{0.15mm}
\put(12,7.5){\line(1,0){1}}
\put(10,7.5){\line(1,0){1}}
\put( 8,7.5){\line(1,0){1}}
\put(7.5,12){\line(0,1){1}}
\put(7.5,10){\line(0,1){1}}
\put(7.5, 8){\line(0,1){1}}
\put(2,7.5){\line(1,0){5.8}}
\put(7.5,2){\line(0,1){5.8}}
\end{picture}  }
\put(0,15){
\parbox[t]{15cm}{
\begin{eqnarray}
\label{allweights}
&&\hspace{-2mm}
\omega_1\hspace{1.375cm}\omega_3\hspace{1.375cm}
\omega_4\hspace{1.375cm}\omega_5\hspace{1.375cm}\omega_6\hspace{1.375cm}
\omega_7\hspace{1.375cm}\omega_8\hspace{3.1cm}\nonumber
\end{eqnarray}  }}
\end{picture}

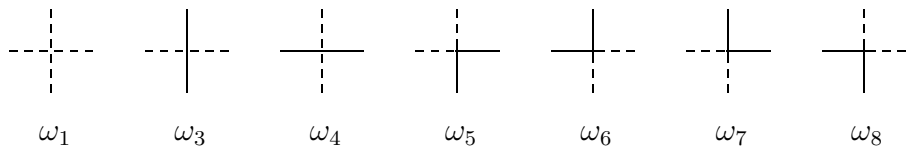
{\bf Figure 1:} Vertices of the (one-colour) self-avoiding loop model
(The intersection vertex is omitted because its weight $\omega_2$ is
zero.). Loop segments are denoted by continuous lines.
\end{table}

\noindent
This model is a special 8-vertex model 
(with $\omega_2 = 0$, i.e.\ a 7-vertex model) 
and, therefore, we have applied here the standard 8-vertex model 
notation (for a discussion of the model also see \cite{scharn2}). 
The vertices of the two-colour
loop model equivalent to the Thirring model can now be found by 
taking two differently coloured sets of the vertices shown in 
Fig.\ 1 and overlapping them in any possible way. This 
leads to 49 two-colour loop model vertices (because
each one-colour loop model has seven vertices with non-vanishing weight).
The weights of 48 out of the total 49 of these two-colour vertices 
are simply given by the product of the weights of the one-colour 
vertices they 
are built of. The only exception is the overlap of the two one-colour
monomer vertices (which is the two-colour monomer vertex) which 
has the weight $M^2 - 2G$ instead of $M^2$ (cf.\ eq.\ (\ref{V1})).\\

\section{Discussion and conclusions}

The value $c = 1$ of the central charge $c$ 
of the one-flavour Thirring model, which is independent 
of the coupling $G$, has been discussed by various methods 
in \cite{dest}, \cite{xue}, \cite{xu}. With the concept 
of universality in mind, this result is not surprising
since a variation of $G$ only affects the monomer weight
of the two-colour loop model equivalent to the Thirring
model, consequently at least in a certain vicinity 
of $G = 0 $ the Thirring model should be expected to exhibit the same
value of the central charge as free Dirac fermions.\\

Let us also comment on the relation of the above two-colour 
loop model to other statistical systems.
Somewhat similar two-colour loop models have recently been 
discussed in \cite{war}. Furthermore, the two-colour 
loop model equivalent to
the Thirring model can also be viewed as a 4-state vertex model.
To see this note that the one-colour self-avoiding loop model,
whose vertices are shown in Fig.\ 1, is a 2-state model. This
leads, by construction, to the understanding of the Thirring model
as a 4-state 49-vertex model. Q-state vertex models have been 
studied, e.g.\ in \cite{strog}-\cite{kash}. However, the 
particular 4-state models discussed in \cite{fate}, 
\cite{akut}, \cite{kash} 
do not contain the case of the 4-state vertex model equivalent to the 
Thirring model which seems to not have been investigated 
in the published literature on the subject
(except, in view of eq.\ (\ref{B15}), the free fermion 
case, where one may use the results obtained in 
\cite{bazh}). 
Finally, it is interesting to note, just as an aside, 
that also certain other models
with four states have recently been discussed within the 
context of fermionic systems in \cite{ord}, \cite{hinr}.\\

The present investigation also immediately allows to 
draw conclusions about another lattice model, namely the one-flavour
strong (infinite) coupling Schwinger model with an additional 
four-fermion interaction term (i.e.\ the Thirring-Schwinger 
model with some finite four-fermion coupling constant
$G$ as in the Thirring model itself).
For $G = 0$ Salmhofer has found the exact equivalence of this model
to a (one-colour) self-avoiding loop model with monomer weight $z = M^2$ and 
bending rigidity $\eta = 1/2$ \cite{salm}. From our consideration
of the Thirring model immediately follows, that for $G \neq 0$
one only has to replace the monomer weight 
$z = M^2$ by $z = M^2 - 2G$ in order to obtain the corresponding
equivalence. Consequently, the model can be discussed
along exactly the same lines as done for $G = 0$ in \cite{scharn2}.\\

To conclude, in the present article we have found the exact equivalence
of the one-flavour lattice Thirring model with Wilson fermions and 
Wilson parameter $r = 1$ to a two-colour loop model which
can also be viewed as a 4-state 49-vertex model on the square
lattice. The choice $r = 1$ has proved crucial in simplifying 
the analysis. It should be emphasized that the two-colour
loop model is a statistical model with positive weights
which allows its numerical simulation by standard methods.
The loop model picture even might prove advantageous because it
allows the application of efficient cluster algorithms
(e.g.\ see \cite{mont1}, \cite{ever}). The particular
loop model equivalence found for free fermions 
(cf.\ eq.\ (\ref{B15})) might be helpful
in the future analysis of multi-flavour four-fermion models
in 2D.  Also, it might serve as a suggestive starting
point for the further analysis of Wilson fermions in higher 
dimensions.\\

\vspace{1.5cm}

\noindent
{\bf Acknowledgements}\\

The present work has been performed under the EC Human 
Capital and Mobility Program, contract no.\ ERBCHBGCT930470
and the EC Training and Mobility of Researchers Program,
return fellowship contract no.\ ERBFMBICT961197.
I would like to thank Simon Hands for discussions and a critical 
reading of the draft version of the paper. Helpful e-mail 
conversations with C.\ Krattenthaler are also gratefully acknowledged.
I am indebted to him for the argument cited in Appendix A.\\

\newpage

\setcounter{section}{1}
\section*{Appendix A}
\renewcommand{\theequation}{\mbox{\Alph{section}.\arabic{equation}}}
\setcounter{equation}{0}

In this Appendix we demonstrate that the relation
\begin{eqnarray}
\label{B10}
C_{-}(L)&\equiv& q(L) + 1\ ({\rm mod}\ 2)
\end{eqnarray} 
holds, where $q(L)$ is the number of self-intersections of a given 
thin line loop $L$ and $C_{-}(L)$ is the number of (oriented)
right-up, up-right segments (cf.\ the second term on the 
r.h.s.\ of eqs.\ (\ref{S3}), (\ref{S4})).
We only consider loops which do not occupy a link twice which, in
principle, is possible due to links with $k_l = 2$ (thick lines).
One may always put such loops which visit at least one link twice
on a finer lattice where each link can be occupied only once. The 
analysis demonstrating the validity of eq.\ (\ref{B10}) 
consequently can also be extended to such cases.\\

We start with $q=0$. The proof will be based on an induction 
with respect to the area $\Omega(L)$ enclosed by the 
loop $L$. If the area $\Omega(L)$
of $L$ is one (in lattice units), i.e.\ $\vert L\vert = 4$
($\vert L\vert$ is the length of the loop),
one may convince oneself that eq.\ (\ref{B10})
holds ($C_{-} = 1$ in this case). Consider now a loop $L$ 
with $\Omega(L) > 1$. There exists at least one loop $L^\prime$
with $\Omega(L^\prime) = \Omega(L) - 1$ whose interior differs from
$L$ by just one lattice square (This selection of $L^\prime$ 
is important because otherwise certain transformations 
to be discussed below would possibly not apply.). 
Consequently, we may perform 
a deformation of the loop $L^\prime$ into the loop $L$.
We will demonstrate that in such a transformation 
$\Delta C_{-} = C_{-}(L^\prime) - C_{-}(L)\ \equiv 0\ ({\rm mod}\ 2)$
holds. In the transformation a lattice square is added to
the interior of the loop $L^\prime$ which has at least one boundary link 
belonging to the loop $L^\prime$. There are three different cases to 
be considered: the lattice square under consideration either
has one ((\ref{trans1})), 
two  ((\ref{trans2}))
or three  ((\ref{trans3}))
boundary links belonging to the loop $L^\prime$
(In the case of two links we do not need to consider the case where
two opposite sites of the lattice square belong to the loop
$L^\prime$ because this would be in contradiction with our choice for 
$L^\prime$.). Below we depict some of the situations that may occur.
The complete set is obtained by rotating the pictures in
(\ref{trans1})-(\ref{trans3}) in all possible ways.

\vspace{5mm}

\noindent
\unitlength1.mm
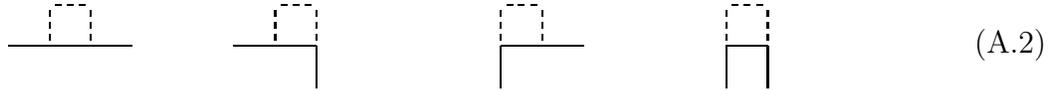
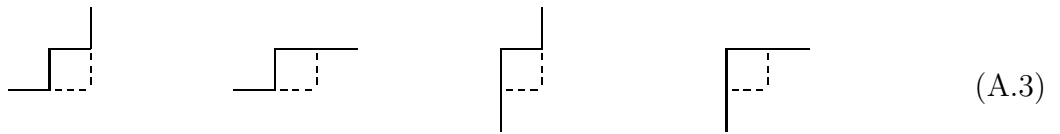
\begin{picture}(150,15)
\put(10,0){
\unitlength1.mm
\begin{picture}(22.5,15)
\linethickness{0.15mm}
\put(2,7.5){\line(1,0){5.5}}
\put(7.5,7.5){\line(1,0){5.5}}
\put(13,7.5){\line(1,0){5.5}}
\put( 11.5,13){\line(1,0){0.78}}
\put( 9.84,13){\line(1,0){0.78}}
\put( 8.28,13){\line(1,0){0.78}}
\put(13,11.5){\line(0,1){0.78}}
\put(13,9.84){\line(0,1){0.78}}
\put(13,8.28){\line(0,1){0.78}}
\put(7.5,11.5){\line(0,1){0.78}}
\put(7.5,9.84){\line(0,1){0.78}}
\put(7.5,8.28){\line(0,1){0.78}}
\end{picture} }
\put(40,0){
\unitlength1.mm
\begin{picture}(22.5,15)
\linethickness{0.15mm}
\put(2,7.5){\line(1,0){5.5}}
\put(7.5,7.5){\line(1,0){5.5}}
\put(13,7.5){\line(0,-1){5.5}}
\put( 11.5,13){\line(1,0){0.78}}
\put( 9.84,13){\line(1,0){0.78}}
\put( 8.28,13){\line(1,0){0.78}}
\put(13,11.5){\line(0,1){0.78}}
\put(13,9.84){\line(0,1){0.78}}
\put(13,8.28){\line(0,1){0.78}}
\put(7.5,11.5){\line(0,1){0.78}}
\put(7.5,9.84){\line(0,1){0.78}}
\put(7.5,8.28){\line(0,1){0.78}}
\end{picture} }
\put(70,0){
\unitlength1.mm
\begin{picture}(22.5,15)
\linethickness{0.15mm}
\put(7.5,7.5){\line(0,-1){5.5}}
\put(7.5,7.5){\line(1,0){5.5}}
\put(13,7.5){\line(1,0){5.5}}
\put( 11.5,13){\line(1,0){0.78}}
\put( 9.84,13){\line(1,0){0.78}}
\put( 8.28,13){\line(1,0){0.78}}
\put(13,11.5){\line(0,1){0.78}}
\put(13,9.84){\line(0,1){0.78}}
\put(13,8.28){\line(0,1){0.78}}
\put(7.5,11.5){\line(0,1){0.78}}
\put(7.5,9.84){\line(0,1){0.78}}
\put(7.5,8.28){\line(0,1){0.78}}
\end{picture} }
\put(100,0){
\unitlength1.mm
\begin{picture}(22.5,15)
\linethickness{0.15mm}
\put(7.5,7.5){\line(0,-1){5.5}}
\put(7.5,7.5){\line(1,0){5.5}}
\put(13,7.5){\line(0,-1){5.5}}
\put( 11.5,13){\line(1,0){0.78}}
\put( 9.84,13){\line(1,0){0.78}}
\put( 8.28,13){\line(1,0){0.78}}
\put(13,11.5){\line(0,1){0.78}}
\put(13,9.84){\line(0,1){0.78}}
\put(13,8.28){\line(0,1){0.78}}
\put(7.5,11.5){\line(0,1){0.78}}
\put(7.5,9.84){\line(0,1){0.78}}
\put(7.5,8.28){\line(0,1){0.78}}
\end{picture} }
\put(0,16){
\parbox[t]{15cm}{
\begin{eqnarray}
\label{trans1}
&&
\end{eqnarray}  }}
\end{picture}
\vspace{2mm}

\noindent
\unitlength1.mm
\begin{picture}(150,22.5)
\put(10,0){
\unitlength1.mm
\begin{picture}(22.5,22.5)
\linethickness{0.15mm}
\put(2,7.5){\line(1,0){5.5}}
\put(7.5,13){\line(1,0){5.5}}
\put( 11.5,7.5){\line(1,0){0.78}}
\put( 9.84,7.5){\line(1,0){0.78}}
\put( 8.28,7.5){\line(1,0){0.78}}
\put(7.5,7.5){\line(0,1){5.5}}
\put(13,13){\line(0,1){5.5}}
\put(13,11.5){\line(0,1){0.78}}
\put(13,9.84){\line(0,1){0.78}}
\put(13,8.28){\line(0,1){0.78}}
\end{picture} }
\put(40,0){
\unitlength1.mm
\begin{picture}(22.5,22.5)
\linethickness{0.15mm}
\put(2,7.5){\line(1,0){5.5}}
\put(7.5,13){\line(1,0){5.5}}
\put( 11.5,7.5){\line(1,0){0.78}}
\put( 9.84,7.5){\line(1,0){0.78}}
\put( 8.28,7.5){\line(1,0){0.78}}
\put(7.5,7.5){\line(0,1){5.5}}
\put(13,13){\line(1,0){5.5}}
\put(13,11.5){\line(0,1){0.78}}
\put(13,9.84){\line(0,1){0.78}}
\put(13,8.28){\line(0,1){0.78}}
\end{picture} }
\put(70,0){
\unitlength1.mm
\begin{picture}(22.5,22.5)
\linethickness{0.15mm}
\put(7.5,7.5){\line(0,-1){5.5}}
\put(7.5,13){\line(1,0){5.5}}
\put( 11.5,7.5){\line(1,0){0.78}}
\put( 9.84,7.5){\line(1,0){0.78}}
\put( 8.28,7.5){\line(1,0){0.78}}
\put(7.5,7.5){\line(0,1){5.5}}
\put(13,13){\line(0,1){5.5}}
\put(13,11.5){\line(0,1){0.78}}
\put(13,9.84){\line(0,1){0.78}}
\put(13,8.28){\line(0,1){0.78}}
\end{picture} }
\put(100,0){
\unitlength1.mm
\begin{picture}(22.5,22.5)
\linethickness{0.15mm}
\put(7.5,7.5){\line(0,-1){5.5}}
\put(7.5,13){\line(1,0){5.5}}
\put( 11.5,7.5){\line(1,0){0.78}}
\put( 9.84,7.5){\line(1,0){0.78}}
\put( 8.28,7.5){\line(1,0){0.78}}
\put(7.5,7.5){\line(0,1){5.5}}
\put(13,13){\line(1,0){5.5}}
\put(13,11.5){\line(0,1){0.78}}
\put(13,9.84){\line(0,1){0.78}}
\put(13,8.28){\line(0,1){0.78}}
\end{picture} }
\put(0,16){
\parbox[t]{15cm}{
\begin{eqnarray}
\label{trans2}
&&
\end{eqnarray}  }}
\end{picture}

\vspace{5mm}

\noindent
\unitlength1.mm
\begin{picture}(150,15)
\put(10,0){
\unitlength1.mm
\begin{picture}(22.5,15)
\linethickness{0.15mm}
\put(2,7.5){\line(1,0){5.5}}
\put(7.5,13){\line(1,0){5.5}}
\put(13,7.5){\line(1,0){5.5}}
\put( 11.5,7.5){\line(1,0){0.78}}
\put( 9.84,7.5){\line(1,0){0.78}}
\put( 8.28,7.5){\line(1,0){0.78}}
\put(7.5,7.5){\line(0,1){5.5}}
\put(13,7.5){\line(0,1){5.5}}
\end{picture} }
\put(40,0){
\unitlength1.mm
\begin{picture}(22.5,15)
\linethickness{0.15mm}
\put(2,7.5){\line(1,0){5.5}}
\put(7.5,13){\line(1,0){5.5}}
\put(13,7.5){\line(0,-1){5.5}}
\put( 11.5,7.5){\line(1,0){0.78}}
\put( 9.84,7.5){\line(1,0){0.78}}
\put( 8.28,7.5){\line(1,0){0.78}}
\put(7.5,7.5){\line(0,1){5.5}}
\put(13,7.5){\line(0,1){5.5}}
\end{picture} }
\put(70,0){
\unitlength1.mm
\begin{picture}(22.5,15)
\linethickness{0.15mm}
\put(7.5,7.5){\line(0,-1){5.5}}
\put(7.5,13){\line(1,0){5.5}}
\put(13,7.5){\line(1,0){5.5}}
\put( 11.5,7.5){\line(1,0){0.78}}
\put( 9.84,7.5){\line(1,0){0.78}}
\put( 8.28,7.5){\line(1,0){0.78}}
\put(7.5,7.5){\line(0,1){5.5}}
\put(13,7.5){\line(0,1){5.5}}
\end{picture} }
\put(100,0){
\unitlength1.mm
\begin{picture}(22.5,15)
\linethickness{0.15mm}
\put(7.5,7.5){\line(0,-1){5.5}}
\put(7.5,13){\line(1,0){5.5}}
\put(13,7.5){\line(0,-1){5.5}}
\put( 11.5,7.5){\line(1,0){0.78}}
\put( 9.84,7.5){\line(1,0){0.78}}
\put( 8.28,7.5){\line(1,0){0.78}}
\put(7.5,7.5){\line(0,1){5.5}}
\put(13,7.5){\line(0,1){5.5}}
\end{picture} }
\put(0,16){
\parbox[t]{15cm}{
\begin{eqnarray}
\label{trans3}
&&
\end{eqnarray}  }}
\end{picture}

\vspace{5mm}

\noindent
One may now provide the loop segments in
(\ref{trans1})-(\ref{trans3}) with an orientation
($L^\prime$ is denoted by a continuous line.).
Then, for each possible case one may explicitly 
convince oneself that in the transition from the loop
$L^\prime$ to the loop $L$ $\Delta C_{-}\in \{-2,0,2\}$. 
According to our induction assumption eq.\ (\ref{B10})
holds for the loop $L^\prime$ ($\Omega(L^\prime) = \Omega(L) - 1$), 
consequently it also holds for $L$. Therefore, eq.\ (\ref{B10}) 
is valid for any loop with $q=0$.
Now we are prepared to discuss the case $q > 0$ (I am indebted 
to C.\ Krattenthaler for the argument \cite{krat}.). We perform 
an induction with respect to $q$. The case $q = 0$ has just
been considered. Next, assume a given oriented loop $L$
has $q(L) > 0$ self-intersections. Select a certain intersection and 
apply to it a reinterpretation in terms of two oriented line 
segments passing each other without intersection. This leads to 
two different loops $A$ and $B$ with $q(A) < q(L) $ and $q(B) < q(L)$ 
intersections each ($q(A) + q(B) = q(L) - 1$). 
As one may convince oneself easily, in this operation 
in three of the four different cases the total number of 
(oriented) right-up, up-right segments does not change. $C_{-}$ only goes up
by 2 in the case of the following intersection diagram.

\parindent0.em
\vspace{5mm}

\unitlength1.mm
\begin{picture}(150,14.5)
\put(3,0){
\unitlength1.mm
\begin{picture}(14.5,14.5)
\linethickness{0.15mm}
\put(13.75,7.25){\line(-1,0){2.25}}
\put(7.25,7.25){\vector(1,0){4.25}}
\put(7.25,7.25){\line(-1,0){2.25}}
\put(0.75,7.25){\vector(1,0){4.25}}
\put(7.25,13.75){\line(0,-1){2.25}}
\put(7.25,7.25){\vector(0,1){4.25}}
\put(7.25,7.25){\line(0,-1){2.25}}
\put(7.25,0.75){\vector(0,1){4.25}}
\end{picture} }
\put(0,16){
\parbox[t]{15cm}{
\begin{eqnarray}
\label{cross}
&&
\end{eqnarray}  }}
\end{picture} 

\vspace{5mm}
Consequently, within the considered operation $C_{-}$ remains 
constant modulo 2.
Now, according to our induction assumption eq.\ (\ref{B10}) is valid
for the loops $A$ and $B$, therefore
\begin{eqnarray}
\label{B11}
C_{-}\ \equiv\ \left[q(A) + 1\right]\ +\ 
\left[q(B) + 1\right]&\equiv&
q(L) + 1\ ({\rm mod}\ 2)\ .\ \ \ \ 
\end{eqnarray}
This proves eq.\ (\ref{B10}).\\

\end{document}